\documentclass[floatfix,twocolumn,aps,prb,showpacs,superscriptaddress]{revtex4}
\usepackage{amssymb}
\usepackage{graphicx}
\usepackage{dcolumn}
\usepackage{bm}
\begin{document}

\title{Effects of an external magnetic field on the gaps and
     quantum corrections in an ordered Heisenberg antiferromagnet with
     Dzyaloshinskii-Moriya anisotropy}
\author{A. L. Chernyshev}
\affiliation{Department of Physics, University of California, Irvine,
  California 92697} 
\date{\today }

\begin{abstract}  
We study the effects of external magnetic field on the properties 
of an ordered Heisenberg antiferromagnet with the Dzyaloshinskii-Moriya (DM)
interaction. 
Using the spin-wave theory quantum correction to the energy, on-site
magnetization, and uniform magnetization are calculated as a
function of the field $H$ and the DM
anisotropy constant $D$. It is shown that the spin-wave excitations
exhibit an unusual field-evolution of the gaps. 
This leads to various non-analytic
dependencies of the quantum corrections on $H$ and $D$. 
It is also demonstrated that, quite generally, the
DM interaction suppresses quantum fluctuations,
thus driving the system to a more classical ground state.
Most of the discussion is devoted to the spin-$S$, 
two-dimensional square lattice antiferromagnet, 
whose  $S=\frac12$ case is closely
realized in 
K$_2$V$_3$O$_8$ where at $H=0$ the DM anisotropy is hidden 
by the easy-axis anisotropy but is revealed in a finite field. 
The theoretical results for the field-dependence of the
spin-excitation gaps in this material are
presented and the implications for other systems are discussed.
\end{abstract}

\maketitle
\section{Introduction}

The studies of  quantum antiferromagnets (AFs) in external
magnetic field attract significant attention because these systems
exhibit a variety of unusual quantum-mechanical phenomena of general
interest. The modern high magnetic field technology, materials'
synthesis, and experimental probes allow precise measurements
in the regime of the field strength comparable to the characteristic
exchange constant of a system. This has made possible the observations
of condensation of triplet excitations in a variety of
chain, ladder, and weakly-coupled dimer
compounds,\cite{Zheludev,S=1,BEC} 
magnetization
plateaux in frustrated magnets,\cite{plato} and other new
effects.\cite{Dender,Kakurai}  It turned out
that in many cases experimental data deviate significantly from the 
theoretical predictions based on the pure isotropic
Heisenberg model in 
external field.\cite{Dender,Sushkov,S=1a} 
Such deviations are due to anisotropies, most notably
the Dzyaloshinskii-Moriya (DM) anisotropy, which are usually small and
often neglected from zero-field considerations. 
Not only such anisotropies can induce qualitatively new effects, but also the
strength of such effects seems to be significantly amplified by an 
external field. In particular, in the one-dimensional (1D) spin-$\frac12$
chains with the DM interaction spin excitation spectrum exhibits a gap 
$\Delta\propto (DH)^{2/3}$ in contrast with the pure Heisenberg case
where the spectrum remains gapless.\cite{Dender,Affleck} 
Thus, while usually treated as an
insignificant contamination of the problem, the DM
interaction can give rise to an interesting phenomena on its own. 

In the case of the 1D spin-chains an unusual field-dependence
of the gap has been first documented experimentally\cite{Dender} 
 and has received a
theoretical explanation soon after that.\cite{Affleck,YuLu} 
For the higher-dimensional cases 
there is a recent theoretical expectation\cite{Affleck,Sato,Mila} 
that the field-dependence of one of the gaps 
should be $\Delta\propto (DH)^{1/2}$. However, such a behavior has not
been observed in any known system yet. There is also a disagreement
between the recent theoretical works\cite{Sato,Mila} in the low-field
regime with the earlier low-field studies of the
Heisenberg$+$DM problem.\cite{Pinkus,footnote1,Trugman}
The latter works predict a non-zero
gap and also weak ferromagnetism\cite{Dzyaloshinskii} in zero field,
while the former do not. 
We will show below that this discrepancy comes from the higher-order
contributions, which yet result in the effects of order ${\cal O}(D)$
in the excitation gap and spin canting angle.
We will also show that in a realistic system the low-field behavior
can be more delicate. The field-dependence of the gap, quite generally,
takes the form $\Delta\propto\sqrt{D(H-H^*)}$, where 
$H^*\sim D$ can be both positive and negative depending on the
system. This clarifies theoretical expectation for 
experiments. 

Furthermore, most of the studies have excluded from consideration the 
quantum effects associated with the DM interaction. 
The generic question is: does the DM
interaction enhance or suppress quantum fluctuations? This is important for
the systems close to the spin-liquid state, such as triangular
lattice AF Cs$_2$CuCl$_4$ where the DM term is
significant,\cite{Coldea,Veillette,footnote2} 
and also
for the frustrated spin systems where quantum effects may help to
select the ground state.\cite{Kotov,Canals}

In this work we study the excitation spectrum and various $T=0$ static
properties of the square lattice, spin-$S$ Heisenberg model with
the DM interaction. This is the simplest case  of a higher-dimensional system
in which generic trends of the mutual effect of the DM interaction,
external field, and quantum fluctuations on the properties of  an
ordered AF can be studied. 
This choice is also motivated, in part, by the recent
experiments in K$_2$V$_3$O$_8$, a 2D, $S=\frac12$ AF, 
which demonstrates an unusual
spin-reorientation transition in a small field.\cite{Lumsden} 
This effect has been
understood as resulting from an interplay of the DM and easy-axis
anisotropies. 

Altogether, the purpose of this work is threefold. First, to
reconcile the earlier low-field and recent high-field theoretical 
studies of the Heisenberg$+$DM$+$field problem. Second, to
study the dependence of the quantum corrections on the DM
interaction. Third, to analyze an experimental system to which these
results can be applied.

We find that the gap in  one of the magnon branches shows the following
behavior when ${\bf H}\perp{\bf D}$: $\Delta_{\pi\pi}=D$ in $H=0$ field,
$\Delta_{\pi\pi}=\sqrt{D}$ 
for $H\simeq 0.7 H_s$, and $\Delta_{\pi\pi}=\sqrt{3}D^{2/3}$ for
$H=H_s$, where $H_s$ is the saturation field and the energy units
$=4SJ$ are used.
These findings are in agreement with the other
works.\cite{Pinkus,Sato,Mila} 
 We also show that, for
the fields $0<H\alt H_s$, the DM interaction leads to the
suppression of the quantum fluctuations. The dependence 
of the quantum corrections on $D$
for various quantities is often non-analytic. It is
only in the regime of the field close to the saturation, $H\sim H_s$,  
where the DM term enhances quantum fluctuations, effectively leading
to a proliferation of the quantum effects into the classical saturated
phase ($H>H_s$). We also show that some quantum corrections  are singular
in the limit of both $H$ and $D$ going to zero, namely
$\lim_{H\rightarrow 0}\lim_{D\rightarrow 0}\neq \lim_{D\rightarrow
  0}\lim_{H\rightarrow 0}$. We demonstrate that K$_2$V$_3$O$_8$ is an
excellent candidate for the observation of the unusual
field-dependence of the gap, characteristic to the 2D and 3D AFs. In
fact, an additional easy-axis anisotropy helps to switch
between the Ising-like and easy-plane behavior in this system, making
it potentially possible to observe a non-analytic $\Delta\propto
\sqrt{H-H^*}$ behavior of the gap.\cite{footnote3}

The paper is organized as follows. 
In Sec. \ref{Q_effects} 
we describe the technical details, derive the spin-excitation spectrum,
 and present the results for the quantum corrections to various quantities 
in different field regimes.
In Sec. \ref{KVO} we give a quantitative discussion of the excitation
spectrum in 
 K$_2$V$_3$O$_8$.
We conclude by Sec. \ref{Conclusions} which contains a brief discussion 
of our results.

\section{Model, spectrum, and quantum corrections}
\label{Q_effects}

As is mentioned in Introduction, we would like to consider the simplest 
possible case of an ordered AF with the DM anisotropy in an external
field, which would allow to investigate the role of quantum corrections in it.
We therefore study the spin-$S$, nearest-neighbor Heisenberg model
on a square  
lattice with the DM interaction:
\begin{eqnarray}
\label{H}
{\cal H}= \sum_{\langle ij \rangle}\bigg(J{\bf S}_i\cdot {\bf S}_j 
+ {\bf D}_{ij}\cdot({\bf S}_i\times {\bf S}_j)\bigg)
-\sum_i {\bf H}\cdot {\bf S}_i, 
\end{eqnarray}
$\langle ij\rangle$ denotes summation over bonds and ${\bf H}$ is the
magnetic field in the energy units ($g\mu_B=1$). 
The direction of the (staggered) DM vector will be 
chosen along the $z$-axis, ${\bf D}_{ij}=(-1)^i(0,0,D)$. 
Aside from being the simplest and the most common\cite{Sato,Veillette} 
choice, this Hamiltonian also corresponds closely
to K$_2$V$_3$O$_8$.\cite{footnote4}  
We would like to note that, generally, this form of the Hamiltonian
is incomplete, because, as noted in
Refs. \onlinecite{Shekhtman1,Shekhtman2,Yildrim},  
the same microscopic processes that generate
the DM cross-product term also yield the higher-order ${\cal O}(D^2)$
terms, which contribute to some observables on equal footing. 
Without going into much details we refer to
the same works\cite{Shekhtman1,Shekhtman2,Yildrim} 
which show that these additional terms can be cast in the
form of an easy-axis anisotropy with the ``easy'' axis along the 
${\bf D}$-vector, also the case realized in K$_2$V$_3$O$_8$. 
Quite generally, the  DM constant $D$ and the easy-axis anisotropy
constant $E$ are not related to each other in a straightforward way
and thus can be treated as phenomenological parameters.
Therefore, in the following
we would like to concentrate on the most commonly used Hamiltonian 
(\ref{H}) which only contains the cross-product DM-term 
and will consider the effect of the easy-axis anisotropy later in the
context of K$_2$V$_3$O$_8$, Sec. \ref{KVO}.
It is convenient to use units $J=1$ in Eq. (\ref{H})
and this convention is used throughout the 
paper unless noted otherwise.

\subsection{Classical limit}
\label{class}

\begin{figure}[t]  
\includegraphics[width=8cm]{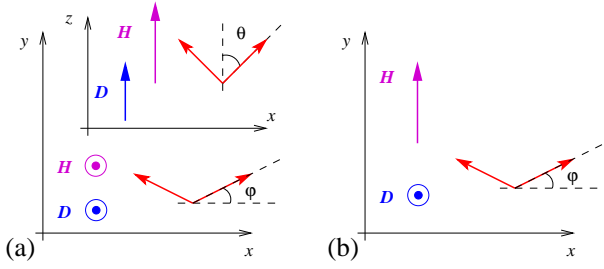}
\caption{(Color online) 
(a) ${\bf H}\| {\bf D}\| z$ configuration. The canting angle
 $\varphi$ in the $x-y$ plane is induced by the DM interaction and does not
  depend on the field. The canting angle $\theta$ towards the field
 ($z$-axis) is given by Eq. (\ref{theta}). (b) ${\bf H}\perp {\bf D}$
 configuration. The canting angle $\varphi$ is given by Eq. (\ref{phi1}).}
\label{angles}
\end{figure}

It is important to discuss the classical limit of the 
model (\ref{H}) 
first, as for the consideration of the excitation spectrum 
the quantization  
axes for spins will be chosen along their classical directions.
The effect of the DM interaction is twofold: it makes energetically
favorable for the  
spins to stay in the plane perpendicular to the direction of ${\bf D}$
and, without  
the $J$-term, it would make the spins in different sublattices to
be under an $\alpha=\pi/2$ 
angle to each other (in the notations of Fig. \ref{angles} 
that corresponds to the angle $\varphi=(\pi-\alpha)/2=\pi/4$). 
While the former trend can only be hindered by additional 
anisotropies (see next Section) the latter always competes with the much 
larger Heisenberg term which prefers the spins to stay antiparallel.
Thus, at $H=0$ in the ground state of the model (\ref{H}) spins lie in the 
$x-y$ plane and are canted towards each other.  
According to Fig. \ref{angles} this configuration has a finite angle
$\varphi(H=0)=\frac12\tan^{-1}D$ 
resulting in a non-zero uniform magnetization $M=S\sin\varphi\simeq
SD/2$ (per spin). 
Therefore, the $H=0$ limit of the model (\ref{H}) 
exhibits an {\it effective} easy-plane anisotropy and a {\it weak
  ferromagnetism}, a phenomenon the DM interaction was originally
proposed for.\cite{Dzyaloshinskii} 

It is clear now that there are two distinct choices for the direction of the 
external field: one is along the DM vector, ${\bf H}\parallel z$, and the 
other one is perpendicular to it, ${\bf H}\perp z$. We will be mostly
concerned with the 
latter case of the in-plane field as it leads to most non-trivial
results and will  
only briefly discuss the former. From the symmetry point of view these
two cases are 
clearly distinct because the symmetry is lowered by the DM interaction
already in zero field from the full rotational $O(3)$ symmetry
to the easy-plane $O(2)$ symmetry. Then the field along the $z$-axis 
will only cant the spins towards itself without affecting the freedom of
choice of the spontaneous in-plane spin direction, thus leaving $O(2)$
intact. On the other hand, the in-plane field will break the remaining 
continuous $O(2)$ symmetry and ``pin'' the spins' direction in the plane.

{\it Canting angles at $D=0$.}\ \ \ 
Since the magnitude of the DM interaction is generally small one may wonder
how much difference its presence 
makes for the ground state of the model (\ref{H}) 
for the finite, and not necessarily small, values of $H$. 
While we will address this issue in detail below it is instructive to consider
the spin configuration when $D=0$. If we neglect the 
DM interaction in (\ref{H}) the only effect of the external magnetic field 
is to induce a finite uniform magnetization by canting 
spins towards the field. The canting angle is $\varphi_0=\sin^{-1}(H/8S)$
and the spins  
become fully polarized in the {\it saturation field} $H_s=8S$ (where
$8=2z$, $z=4$ is  
the coordination number for the square lattice).

\subsubsection{Out-of-plane field.}
We now consider the field-dependence of the canting angles in the
presence of the  
DM interaction. If the field is perpendicular to the $x-y$ plane 
the canting induced by the $D$-term and the one induced by the field lie in the
orthogonal planes, see Fig. \ref{angles}(a). Therefore, one should
expect the DM-induced in-plane canting angle $\varphi$ to be 
independent of the field.
Using the angle notations of Fig. \ref{angles}(a) the classical energy
of the model Eq. (\ref{H}) can be written as:
\begin{eqnarray}
\label{E1}
\frac{E_{cl}}{2NS^2}=1-\sin^2\theta\left(2\cos^2\varphi
+D\sin 2\varphi\right)-\frac{H}{2S}\cos\theta\ ,
\end{eqnarray}
minimizing which gives the canting angles, $N$ is the number of lattice sites.
As expected, $\varphi=\frac12\tan^{-1}D$ is  
independent of the field and the field-induced out-of-plane canting
angle should be found from:
\begin{eqnarray}
\label{theta}
\cos\theta=\frac{H}{8S}\ \frac{2}{1+\sqrt{1+D^2}}\ . 
\end{eqnarray}
Thus, the only effect of the small DM interaction for the case of
${\bf H}\|{\bf D}$  
is the slight change of the saturation field from $H_s=8S$ to 
$H_s^\|=4S\left(1+\sqrt{1+D^2}\right)$.
\begin{figure}[t]  
\includegraphics[width=8cm]{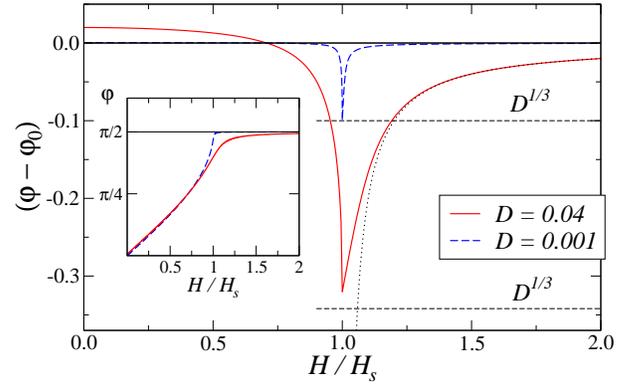}
\caption{(Color online) 
The difference of the in-plane canting angle $\varphi$ and
  its $D=0$ limit $\varphi_0$ v.s. field in the ${\bf H}\perp{\bf D}$
configuration for two representative values of $D=0.001$ (dashed lines) 
and $D=0.04$ (solid lines), $D$ is in units of $J$. 
Inset: the canting angle $\varphi$ v.s. field. The
  asymptotic behavior of $\varphi-\varphi_0=-DH_s/2(H-H_s)$ is 
shown by the dotted line. $D^{1/3}$ values for both choices of $D$ 
are shown by the horizontal dashed lines.}
\label{delta_phi}
\end{figure} 

\subsubsection{In-plane field.}
For the in-plane field the changes for the canting angle are less trivial. 
The spins are in the $x-y$ plane, see Fig. \ref{angles}(b).
From minimizing the classical energy:
\begin{eqnarray}
\label{E2}
\frac{E_{cl}}{2NS^2}=-\cos 2\varphi-D\sin
2\varphi-\frac{H}{2S}\sin\varphi\ ,
\end{eqnarray}
the canting angle obeys the following equation: 
\begin{eqnarray}
\label{phi1}
\sin\varphi-\frac{D}{2}\frac{\cos 2\varphi}{\cos\varphi}=\frac{H}{8S} . 
\end{eqnarray}
Without the DM interaction Eq. (\ref{phi1}) yields
a familiar $H$-dependence of the canting angle: $\varphi_0=\sin^{-1}(H/8S)$
(0 in the subscript refers to $D=0$).
To demonstrate the effect of the DM interaction on the canting
angle in the ${\bf H}\perp{\bf D}$ case 
we plot the difference of $(\varphi-\varphi_0)$ 
v.s. field for two representative values of
$D$ in Fig. \ref{delta_phi}.  
One can see that the difference is $D/2$ for small
field and it vanishes for the field $H\simeq 0.707 H_s$. This field
makes the angle between the spins to be $\pi/4$ for {\it any} value of $D$.
Thus, the DM-term in the energy is always fully minimized for such
a field.
The most striking feature in this plot is the sharp singularity at
$H=H_s$. This is due to the fact that the $D=0$ canting angle
$\varphi_0$  exhibits a kink at $H_s$ where it reaches the maximal
value $\pi/2$. In contrast, the
field-dependence of the canting angle $\varphi$ is smooth through the
saturation field, as shown in the inset of Fig. \ref{delta_phi}. 
In fact, the spins never
fully saturate, only as $H\rightarrow\infty$. The asymptotic behavior
of the angle is: $\varphi-\pi/2=-DH_s/2(H-H_s)$, which is shown in
Fig. \ref{delta_phi} by the dotted line.
Therefore, the
saturation field formally ceases to exist for the ${\bf H}\perp{\bf
  D}$ case. In the rest of the paper we refer to the ``bare'' 
($D=0$) value $H_s=8S$ as to the value of the saturation field.
It is interesting to note that
the magnitude of the deviation of the canting angle 
from its $D=0$ value is quite significant close to the saturation
field. Taking $H=H_s$ and assuming $D\ll 1$ 
one can find from Eq. (\ref{phi1}) that $\cos\varphi\simeq
\pi/2-\varphi\simeq D^{1/3}$, a
value which is not necessarily small even if $D$ itself is reasonably small. 
The $D^{1/3}$ values for the choices of $D$ 
are shown in Fig. \ref{delta_phi} by the 
horizontal dashed lines.

\subsection{Spectrum and gaps}
\label{gaps}

{\it Technical approach.}\ \ \ To apply the spin-wave theory to the
model (\ref{H}) we follow the general prescription of
Ref. \onlinecite{Zhitomirsky}. Namely, the rotating {\it local}
reference frame  for each sublattice is introduced in which the
quantization axis $z_0$ is directed along the {\it classical} spin
direction. In such a local frame the standard bosonic representation
for the spin operators is used:
\begin{eqnarray}
\label{S1}
&&S^{z_0}=S-a^\dag a, \nonumber\\
&&S^{x_0}=\frac{S^+ +S^-}{2}=\sqrt{\frac{S}{2}}(a+a^\dag),\\
&&S^{y_0}=\frac{S^+ -S^-}{2i}=-i\sqrt{\frac{S}{2}}(a-a^\dag),\nonumber
\end{eqnarray}
where we omit higher-order boson terms from $S^\pm$ thus restricting
ourselves to the linear spin-wave approximation. Higher-order, as usual, 
means  
both the higher number of boson operators and the higher powers of $1/S$.
Then one needs a
transformation of the spin components from the local reference frame
to the laboratory reference frame. Such transformations are again 
based on the classical considerations of the canting angles similar to
the ones  
given in Sec. \ref{class}, and they depend on the relative direction of the
field, anisotropies, etc. To give a specific example we provide here 
such a transformation for the ${\bf H}\perp {\bf D}$ configuration for
the vectors ${\bf D}\| z$ and ${\bf H}\| y$, see Fig. \ref{angles}(b):
\begin{eqnarray}
\label{S2}
&&S^x_i = -S^{x_0}_i \sin\varphi +S^{z_0}_i (-1)^i\cos\varphi  \nonumber \\
&&S^y_i = S^{z_0}_i \sin\varphi +S^{x_0}_i (-1)^i\cos\varphi   \\
&&S^z_i = S^{y_0}_i \nonumber \ .
\end{eqnarray}
Using relations of that type and representation
(\ref{S1}) one can rewrite the Hamiltonian (\ref{H}) in the form:
\begin{eqnarray}
\label{H1}
{\cal H}=E_{cl}+{\cal H}_{LSWT}+\dots\ ,
\end{eqnarray}
where $E_{cl}$ is the classical energy discussed above, the 
ellipsis stands for the higher-order terms, 
and the spin-wave Hamiltonian can be written in the general form:
\begin{eqnarray}
\label{H2}
&&{\cal H}_{LSWT}=4S\sum_{\bf k}\bigg(\Big(C_1+C_2
\gamma_{\bf k}\Big)a_{\bf k}^\dag a_{\bf k}^{\phantom \dag} \\
&&\phantom{{\cal H}_{LSWT}=4SJ\sum_{\bf k}\bigg(}
+\frac{C_3\gamma_{\bf k}}{2} 
\left(a_{\bf k}^\dag a_{-{\bf k}}^{\dag}+a_{\bf k}a_{-{\bf k}}\right)
\bigg)\ ,\nonumber
\end{eqnarray}
where $\gamma_{\bf k}=(\cos k_x+\cos k_y)/2$ and the summation over
${\bf k}$ is over the full Brillouin zone (BZ). 
All the coefficients $C_i$
are the functions of the field and anisotropies.
As is noted in  Ref. \onlinecite{Zhitomirsky}, the minimization of the
classical energy ensures the absence of the terms linear in  $a$, or
$a^\dag$ in the 
expansion of the Hamiltonian (\ref{H1}).
Then, the spin-wave Hamiltonian is readily diagonalized by the Bogolyubov
transformation and takes the form:
\begin{eqnarray}
\label{H3}
{\cal H}_{LSWT}=\delta E +\sum_{\bf k}\omega_{\bf k}
\alpha_{\bf k}^\dag \alpha_{\bf k}^{\phantom \dag} \ ,
\end{eqnarray}
where the magnon energy is
\begin{eqnarray}
\label{w}
\omega_{\bf k}=4S\sqrt{\Big(C_1+C_2\gamma_{\bf k}\Big)^2
-C_3^2\gamma^2_{\bf k}}\ ,
\end{eqnarray}
 and
\begin{eqnarray}
\label{dE}
\delta E= \frac12\sum_{\bf k}\Big(\omega_{\bf k}-4SC_1\Big)\ .
\end{eqnarray}
As an example of the use of such a procedure, one can neglect the
DM-term for a moment and, after some algebra, obtain: $C_1=1$,
$C_{2,3}=(-\cos 2\varphi_0 \pm 1)/2$. This leads to the following
expression for the spin-wave spectrum:
\begin{eqnarray}
\label{w0}
\omega_{\bf k}^{D=0}
=4S\sqrt{(1+\gamma_{\bf k})(1-\cos 2\varphi_0\gamma_{\bf k})}\ ,
\end{eqnarray}
where $\cos 2\varphi_0=1-2(H/H_s)^2$, 
in agreement with the results of Ref. \onlinecite{Zhitomirsky}.
\begin{figure}[t]  
\includegraphics[width=7cm]{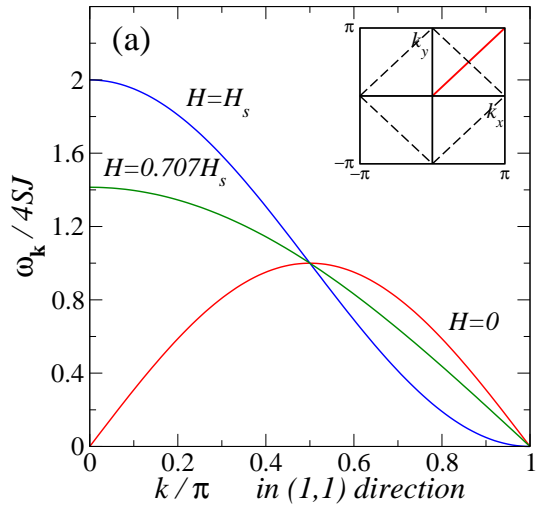}
\caption{(Color online) 
Linear spin-wave result, Eq. (\ref{w0}), for the magnon spectrum
in a field ($D=0$), for $H=0$, $H=H_s/\sqrt{2}$, and $H=H_s$ along the 
main diagonal of the BZ. Inset: full BZ with the direction of the cut.
Magnetic BZ is shown by the dashed diamond.}
\label{wk}
\end{figure} 

{\it General features of the spectrum.} \ \ \
Before we discuss the details of the gaps' dependence on both the
field and the DM interaction, we would like to outline the broad
features of the spin-wave spectrum. 
Our Fig. \ref{wk} shows the $D=0$ spectrum given by Eq. (\ref{w0})
along the main diagonal of the Brillouin zone (see inset) for
three representative values of the field. 
In the $H=0$ case the spectrum has two gapless modes, at $(0,0)$ and
$(\pi,\pi)$. In the field one of the modes
develops a gap, $\Delta_{00}\equiv H$, which is strictly equal to the
value of the field. This mode corresponds to the uniform precession of the
field-induced magnetization about the field direction.\cite{ZhCh} The
other mode must remain gapless as it corresponds to the Goldstone mode
related to the spontaneous breaking of the remaining $O(2)$ symmetry of the
spins in the plane perpendicular to the field.

Two remarks are in
order. First, while it is convenient to use the ``full'' BZ
language here with a single magnon branch, 
one can employ an equivalent picture of two magnon
branches within the magnetic BZ (shown in the inset of
Fig. \ref{wk} by the dashed diamond).
The neutron scattering will observe these branches near the $(\pi,\pi)$-point
in the different components of the structure factor: ${\cal S}^{\pm}\propto
\delta(\omega-\omega_{\bf k})$ and ${\cal S}^{zz}\propto
\delta(\omega-\omega_{\bf k-(\pi,\pi)})$, see Ref. \onlinecite{ZhCh}.
Second, we neglect the higher-order non-linear terms in the spin-wave
Hamiltonian. While they do not change most of the results
qualitatively, in the fields close to the saturation they induce an 
instability of the single-magnon branch towards decays
into a two-magnon continuum.\cite{ZhCh} Since  such an instability does not
concern the ${\bf k}=(0,0)$ and ${\bf k}=(\pi,\pi)$ modes we do not
take this effect into account in this work.\cite{footnote5} 
\begin{figure}[t]  
\includegraphics[width=7cm]{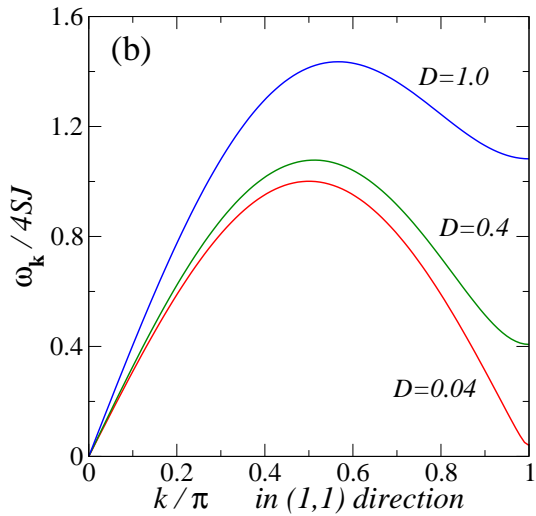}
\caption{(Color online) 
Linear spin-wave result, Eq. (\ref{wD0}), for the magnon spectrum
in zero field for $D=0.04$, $D=0.4$, and $D=1.0$ along the 
main diagonal of the BZ. }
\label{wk_DM}
\end{figure} 

Another general feature of the spin-wave spectrum behavior can be
studied by considering the case $H=0$ and $D\neq 0$.
Using Eqs. (\ref{phi1}), (\ref{S2}), and (\ref{H2}) one obtains
$C_1=\sqrt{1+D^2}$, $C_{2,3}=(-\sqrt{1+D^2}\pm 1)/2$, which gives:
\begin{eqnarray}
\label{wD0}
\omega_{\bf k}^{H=0}
=4S\sqrt{1+D^2}\ \sqrt{(1-\gamma_{\bf k})
\left(1+\frac{\gamma_{\bf k}}
{\sqrt{1+D^2}}\right)}\ .
\end{eqnarray}
This spectrum is plotted in Fig. \ref{wk_DM} for three different
values of $D$. Again, one of the modes is gapless due to 
the $O(2)$ symmetry, which is now related to the spontaneous choice of the
direction of the DM-induced {\it ferromagnetic} magnetization vector. 
The gap in the other
mode is $\Delta_{\pi\pi}\simeq 4S D$ for small $D$. These results are
very similar to the ones obtained in Ref. \onlinecite{Trugman} in the 
context of La$_2$CuO$_4$.

\subsubsection{Out-of-plane field.}
Having outlined the general trends of the spin-wave spectrum behavior
we return to the original problem of the DM interaction in a field. 
We first consider ${\bf H}\|{\bf D}\| z$ configuration. Recalling our
discussion about the canting angles for this case in
Sec. \ref{class}.1, and using 
Eq. (\ref{theta}), after some algebra one obtains:
\begin{eqnarray}
\label{Cpar}
&&C_1=C_2+C_3=\sqrt{1+D^2}\ ,\\
&&C_2-C_3=-1+\frac{4(H/H_s)^2}{1+\sqrt{1+D^2}}\ .\nonumber
\end{eqnarray}
This gives the following spectrum:
\begin{eqnarray}
\label{wpar}
&&\omega_{\bf k}^\|
=4S\sqrt{1+D^2}\ \sqrt{(1+\gamma_{\bf k})
(1-\gamma_{\bf k}\delta)} \ ,\\
&&\mbox{with}\ \ 
\delta=\left(1-\frac{4(H/H_s)^2}{1+\sqrt{1+D^2}}\right)
\frac{1}{\sqrt{1+D^2}}.\nonumber 
\end{eqnarray}
As is discussed in Sec. \ref{class}, this configuration of the field
and the DM vector  
preserves the $O(2)$ symmetry of the problem. Because of that the spectrum
remains essentially identical to the field-only, $D=0$ case, Fig. \ref{wk} 
and Eq. (\ref{w0}). That is, the $(\pi,\pi)$-mode is gapless and the 
$(0,0)$-mode is gapped with $\Delta_{00}=4S\sqrt{D^2+4(H/H_s)^2}$.
The other changes concern the saturation field, which is increased to  
$H_s^\|=4S\left(1+\sqrt{1+D^2}\right)$, and 
 the total width of the spectrum, which is increased by the factor
$\sqrt{1+D^2}$.  
With these results we conclude our consideration of the ${\bf H}\|{\bf
  D}$ case. 

\subsubsection{In-plane field.}
We finally arrive to the most interesting part of the consideration: 
the orthogonal configuration of the field and the DM vector, 
${\bf H}\perp {\bf D}\|z$. 
As we have seen in Sec. \ref{class}.2, this case is 
characterized by the absence of the true saturation of the spins and by
an intriguing behavior of the canting angle. As is already mentioned, the 
non-zero DM constant and external field break all continuous
symmetries and thus no excitation branch should remain gapless. 
However, since the symmetry-breaking effect due to a small DM
interaction should be weak one expects the gap associated with it to 
be small as well.

With these ideas in mind one can use Fig. \ref{angles}(b), expression for
the canting angle for this case, Eq. (\ref{phi1}), transformation
from the local to the laboratory frame, Eq. (\ref{S2}), and
the transformation of Eq. (\ref{H}) to Eq. (\ref{H3}) to obtain:
\begin{eqnarray}
\label{Cperp}
&&C_1=1+D\tan\varphi\ , \nonumber\\
&&C_2+C_3=-\cos 2\varphi - D\sin 2\varphi\ ,\\
&&C_2-C_3=1\ , \nonumber
\end{eqnarray}
where the canting angle is defined from Eq. (\ref{phi1}).
This leads to the following expression for the spin-wave dispersion:
\begin{eqnarray}
\label{wperp}
\omega_{\bf k}^\perp =4S\sqrt{(C_1+\gamma_{\bf k})
\left(C_1-\left(C_1-\frac{H}{4S}\sin \varphi\right)\gamma_{\bf k}\right)} \ .
\end{eqnarray}
Just to verify the consistency of this result with our previous
considerations one can set $D=0$ and obtain from Eq. (\ref{phi1}): $C_1=1$ and
$\sin\varphi=H/8S$. This yields the result for
the Heisenberg$+$field spectrum obtained in Eq. (\ref{w0}) and shown in
Fig. \ref{wk}. On the other hand, setting field to zero yields 
$\tan2\varphi=D$ from Eq. (\ref{phi1}) and, after
some algebra, one reproduces the $H=0$ spectrum for the
Heisenberg$+$DM problem given in Eq. (\ref{wD0}) and shown in
Fig. \ref{wk_DM}. 

\begin{figure}[t]  
\includegraphics[width=8cm]{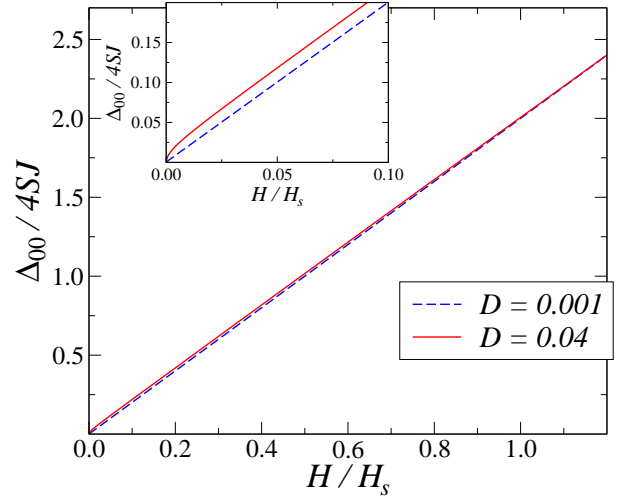}
\caption{(Color online) 
$\Delta_{00}/4SJ$ v.s. $H/H_s$ for $D=0.001$ and
  $D=0.04$. Inset: same plot in the region of small fields.}
\label{gapsA00}
\end{figure}

{\it $(0,0)$-gap.} \ \ \
Given the above discussion, the broad features of the spectrum
$\omega_{\bf k}^\perp$ should look like a convolution of the Figures
\ref{wk} and \ref{wk_DM}. Namely, it should have gaps at both $(0,0)$
and $(\pi,\pi)$ points that evolve with field. 
We first study such an evolution for the gap at ${\bf k}=(0,0)$, which
is given by:
\begin{eqnarray}
\label{gap1}
\Delta_{00}=4S\sqrt{\frac{H}{2S}\sin\varphi
\left(1+\frac12D\tan\varphi\right)} \ ,
\end{eqnarray}
with $H$- and $D$-dependence of $\varphi$ to be defined from
Eq. (\ref{phi1}). This gap is plotted as a function of field in
Fig. \ref{gapsA00} for two representative values of $D$. We recall
that in the $D=0$ case $\Delta_{00}$ grows linearly with $H$ with a
slope equal to unity. Since we plot $\Delta_{00}/4S$ v.s. $H/H_s$ with
$H_s=8S$, this translates into the slope $=2$ in these units. One can
see that for the small values of $D$ the difference of our results
from $D=0$ case mainly concerns small fields $H\ll H_s$. In this
limit, assuming $D\ll 1$, one can simplify Eqs. (\ref{gap1}) and
(\ref{phi1}) to obtain:
\begin{eqnarray}
\label{gap1a}
&&\Delta_{00}\simeq 8S\sqrt{\frac{H}{H_s}
\left(\frac{H}{H_s}+\frac12 D\right)}\\
&&\phantom{\frac{\Delta_{00}}{4S}}
\simeq
\left\{
\begin{array}{ll}
\sqrt{4SD}\,\sqrt{H}&\ \ \mbox{for} \ \  H\ll D\ ,\\ \\
H+2SD&\ \ \mbox{for} \ \  D\ll H\ll H_s\ .
\end{array}
\right. \nonumber
\end{eqnarray}
Thus, for very small fields this gap is $\propto \sqrt{H}$ and for
$H\agt D$ it is a linear function of $H$ with an offset $\propto
D$. This behavior can be seen in the inset of Fig. \ref{gapsA00} which
zooms into the region of small fields. 
Altogether, the changes from the DM-term 
in the field-dependence of the ${\bf k}=(0,0)$
gap are small and are due to the fact that the DM interaction induces a
non-zero canting of spins already in zero field.

\begin{figure}[t]  
\includegraphics[width=8cm]{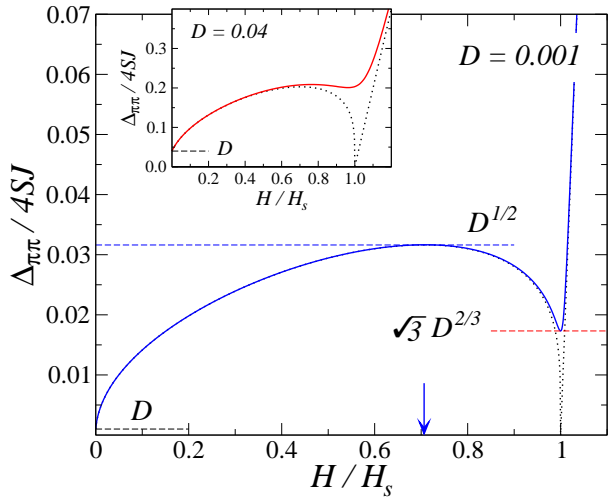}
\caption{(Color online) 
Field dependence of $\Delta_{\pi\pi}/4SJ$ v.s. $H/H_s$ for
  $D=0.001$, Eq. (\ref{gap2}). 
The gap values at $H=0$, $H=H_s/\sqrt{2}$, and $H=H_s$
  are marked by the horizontal dashed lines.
Dotted lines are: for $H< H_s$ -- the low- and intermediate-field
  approximations for the gap, Eq. (\ref{gap2a}), for   $H> H_s$ -- the
  $D=0$ gap dependence $\Delta_{\pi\pi}=H-H_s$. 
Arrows mark $H=H_s/\sqrt{2}$. Inset: same for $D=0.04$.}
\label{gapsA}
\end{figure}

{\it $(\pi,\pi)$-gap.} \ \ \
The evolution of the ${\bf k}=(\pi,\pi)$ gap is much more interesting,
because it corresponds to a mode which would be gapless in $D=0$
case. Using Eqs. (\ref{wperp}) one arrives to:
\begin{eqnarray}
\label{gap2}
\Delta_{\pi\pi}=4S\sqrt{2D\tan\varphi
\left(1+D\tan\varphi-\frac{H}{8S}\sin\varphi\right)} \ ,
\end{eqnarray}
which clearly vanishes as $D\rightarrow 0$. The angle
$\varphi$ is to be defined from Eq. (\ref{phi1}) as before.
Considering $D\ll 1$ limit one can identify three different regimes
for the field: $H\sim D$, $H_s\agt H\gg D$, and $H\simeq H_s$.
After some algebra one can simplify Eq. (\ref{gap2}) 
and obtain the following expressions for the gap
in these regimes:
\begin{eqnarray}
\label{gap2a}
\Delta_{\pi\pi}\simeq 
\left\{
\begin{array}{ll}
\sqrt{4SD}\,\sqrt{H+4SD}&\ \ \mbox{for} \ \  H\sim D\ ,\\ \\
\sqrt{4SD}\,\sqrt{H\sqrt{1-\left(\displaystyle\frac{H}{8S}\right)^2}}&\ \ 
\mbox{for} \ \  H\gg D \ ,\\ \\
4S\, \sqrt{3}\, D^{2/3}&\ \ \mbox{for} \ \  H = H_s\ .
\end{array}
\right. 
\end{eqnarray}
These trends are clearly seen in our Fig. \ref{gapsA} for a small
value of $D=0.001$. The 
gap evolves from $\Delta_{\pi\pi}/4S=D$ in zero field to its maximal value 
$\Delta_{\pi\pi}/4S=D^{1/2}$ at $H\simeq 0.707 H_s$, 
and then has a minimum at $H_s$ where it reaches the value
$\sqrt{3}D^{2/3}$ (marked by the dashed horizontal lines in
Fig. \ref{gapsA}).  
The origin of the maximum can be
simply understood from Eq. (\ref{gap2}) assuming $H\gg D$ and
neglecting $D$ from everywhere except the common prefactor. This
gives another expression for the $(\pi,\pi)$-gap in the
intermediate-field regime:
\begin{eqnarray}
\label{gap2b}
\Delta_{\pi\pi}\simeq 4S\sqrt{D\sin 2\varphi}, \ \
\mbox{with} \ \ \sin\varphi\approx\frac{H}{8S}\ .
\end{eqnarray}
This expression obviously has a maximum at $\varphi=\pi/4$
($H=H_s/\sqrt{2}$), the canting angle at which the DM interaction is
fully satisfied. 

One can see that for the higher value of $D=0.04$ the
maximum and minimum in the field-dependence of the gap are only weakly
pronounced (inset of Fig. \ref{gapsA}), while the zero-field gap
$=D$
persists. Note that the zero-field gap is  a salient feature of the
model (\ref{H}), intimately related to the phenomenon of the 
weak ferromagnetism.
The origin of the minimum and of the unusual $2/3$ power of $D$ 
can be traced to the classical consideration we gave for the
canting angle at the saturation field in Se. \ref{class}.2. 
At $H=H_s$ we have
$\cos\varphi\simeq D^{1/3}$, which yields $\Delta_{\pi\pi}\propto
D^{2/3}$ from Eq. (\ref{gap2}). For an interesting discussion of that
fractional power see Ref. \onlinecite{Mila}.

The minimum in $\Delta_{\pi\pi}(H)$ disappears, or, 
more precisely, merges with the maximum and
becomes an inflection point, at a somewhat higher value of
$D_c\simeq 0.066$. A similarly small number has been obtained  in
Ref. \onlinecite{Mila} in the mean-field study of the Lieb-Mattis
model.\cite{footnote6} It is surprising to have such a small
``magic number'' and one may wonder whether there is some hidden small
parameter behind it. It turns out that it is a small 
difference of the powers of $D$ in  $\Delta^{max}_{\pi\pi}$ and
$\Delta^{min}_{\pi\pi}$ which is responsible for that smallness. 
One can estimate the ``critical'' value of $D$ at which the minimum
disappears as the point where
$\Delta^{max}_{\pi\pi}\approx\Delta^{min}_{\pi\pi}$. This gives:
$D_c^{1/6}=1/\sqrt{3}$, or $D_c=1/27\simeq 0.037$. The actual number is
about two times higher, but already at $D=0.04$ the minimum has almost
faded away, see Fig. \ref{gapsA}.

To conclude the description of Fig. \ref{gapsA}, the dotted line for
$H<H_s$ is the convolution of the low- and intermediate-field results
from Eq. (\ref{gap2a}), which gives a very close description for the
gap behavior up to $H\sim H_s$.  The dotted line for $H>H_s$ is the
$D=0$ field dependence of the gap above the saturation field: 
$\Delta_{\pi\pi}=H-H_s$. One can
see that the gap merges with this asymptote quickly. We add that 
the results similar to that in Fig. \ref{gapsA}
were also reported for a 1D model with the DM interaction in
Ref. \onlinecite{YuLu}. 

Our result for $\Delta_{00}$ and $\Delta_{\pi\pi}$ in the low-field
regime, Eqs. (\ref{gap1a}) and (\ref{gap2a}), 
are in agreement with the results of
the earlier works from the 1960s.\cite{Pinkus} 
For the intermediate and high-field regimes our results in
Eq. (\ref{gap2a}) agree with the results of the recent works on the
coupled chains\cite{Sato} and Lieb-Mattis model.\cite{Mila} These
works used the effective staggered field in place of the DM
interaction that allows to simplify the model but introduces an
approximation that neglects effects of order ${\cal O}(D)$ in the
gaps.  Such an approximation 
has created a discrepancy of these works with the
earlier studies in the low-field regime. Namely, these recent 
works predict
no weak ferromagnetism and no gap for the $(\pi,\pi)$-mode in zero
field.
Our results for $\Delta_{\pi\pi}$, Eqs. (\ref{gap2}) and
(\ref{phi1}), provide a reconciliation of 
the earlier low-field and recent high-field studies. We also 
have provided a detailed analysis of the gap in the other spin-wave
branch, the $(0,0)$-mode, Eqs. (\ref{gap1}) and (\ref{gap1a}), 
which {\it is} gapless in zero field and has a
$\propto\sqrt{H}$ behavior in a  small field regime that can be mistaken
with the results of Refs. \onlinecite{Sato,Mila} for the $(\pi,\pi)$-mode.

This unified description of the behavior of the gaps 
 in external field is also important as it clarifies
the experimental prediction for the gaps' dependence on the
field. In the systems with the DM anisotropy which can be described by
 Eq. (\ref{H}), 
instead of the literal $\propto \sqrt{DH}$ non-analytic
behavior of the $(\pi,\pi)$-gap one
should expect a non-analytic increase from a {\it finite} gap, 
$\propto \sqrt{D(H+4SD)}$,
 according to Eq. (\ref{gap2a}).
As we will show in Sec. \ref{KVO}, in real systems 
this behavior is further altered by the presence of 
other anisotropies.

\subsection{Quantum corrections.}
\label{quant}

We gave an extensive discussion of the $(\pi,\pi)$-gap in
Sec. \ref{gaps} because it will be also
responsible for many of the changes 
in the behavior of quantum corrections  due to the DM interaction.

As is mentioned in Introduction, very little systematic discussion of
the influence of the DM interaction on the quantum effects in the
Heisenberg-like systems has been given in the literature. The recent
exception, Ref. \onlinecite{Veillette}, considers the
triangular-lattice AF$+$DM$+$field, to model the existing experiments
in Cs$_2$CuCl$_4$. However, the results for that system 
are severely complicated by the other effects such as the
field-dependence of the AF ordering vector and are hard to generalize
to the other systems. Considering a much simpler, non-frustrated AF,
we would like to ask a simple question: whether the quantum
fluctuations are enhanced or suppressed by the DM interaction.

{\it Formulae.}\ \ \ We are going to answer this question by analyzing the
 field-dependencies of the $1/S$-corrections to the ground-state
 energy, $\delta E$, local (on-site) magnetization, $\delta S$, and the
 uniform magnetization $\delta M$. As is discussed in Sec. \ref{gaps}
the {\it out-of-plane} field case, ${\bf H}\|{\bf D}\| z$, is almost
 identical to the $D=0$ case, considered in
 Ref. \onlinecite{Zhitomirsky}. Because of that, we are
going to consider  only the case of the {\it in-plane} field, ${\bf
 H}\perp{\bf D}\| z$. 
\begin{figure}[t]  
\includegraphics[width=8cm]{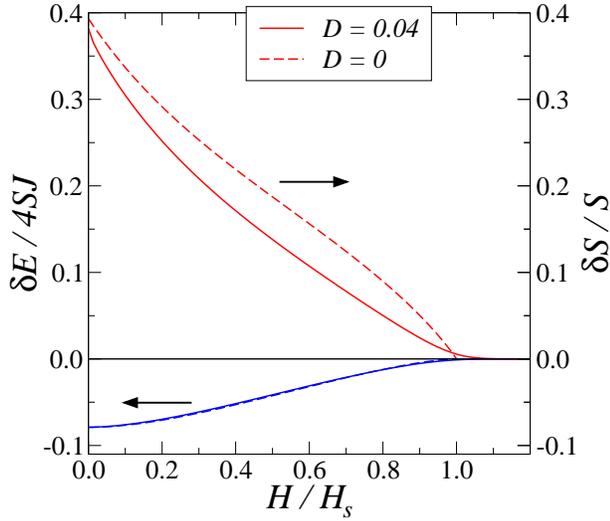}
\caption{(Color online) 
$1/S$-corrections to the ground-state energy (left axis) and
the on-site magnetization (right axis) for $D=0$ (dashed) and $D=0.04$ 
(solid) v.s. $H/H_s$, $S=1/2$.}
\label{q_dE_dS}
\end{figure} 

The $1/S$-correction to the ground-state energy (per spin) is given in
Eq. (\ref{dE}),  which we rewrite here 
\begin{eqnarray}
\label{dEa}
\delta E= 2S\sum_{\bf k}\Big(\nu_{\bf k}-C_1\Big)\ ,
\end{eqnarray}
using the dimensionless spin-wave energy $\nu_{\bf k}\equiv \omega_{\bf
  k}^\perp/4S$, Eq. (\ref{wperp}), normalized to $4S$ to keep the
  $S$-proportionality explicit. 
The definition of the $1/S$-correction to the local (on-site) magnetization
 $\delta S$ follows straightforwardly from:
\begin{eqnarray}
\label{dS}
&&\langle S^{z_0}\rangle = S-\langle a^\dag a\rangle= S-\delta S\
,\nonumber\\ 
&&\delta S =\langle a^\dag a\rangle=\frac12\sum_{\bf
  k}\left(\frac{C_1+C_2\gamma_{\bf k}}{\nu_{\bf k}}-1\right) \ ,
\end{eqnarray}
where the field- and DM-dependent constants $C_i$ are listed in
Eq. (\ref{Cperp}). The classical part of the per-spin 
uniform magnetization is simply given by the canting angle: 
$M_{cl}=S\sin\varphi$, see Fig. \ref{angles}(b). 
A more rigorous definition:
\begin{eqnarray}
\label{Mdef}
M=-\frac{d E}{d H}\ ,
\end{eqnarray}
where $E$ is the energy per spin from Eq. (\ref{E2}), 
leads to the same answer for the
classical part and also yields the $1/S$-correction $\delta M$ as follows:
\begin{eqnarray}
\label{dM}
&&M = M_{cl}-\delta M =S\sin\varphi -\delta M \ 
,\nonumber\\ 
&&\delta M =2S\varphi^\prime\left[\sum_{\bf k}\frac{\partial \nu_{\bf
      k}}{\partial \varphi}-\frac{\partial C_1}{\partial
    \varphi}\right] \ ,
\end{eqnarray}
where
\begin{eqnarray}
\label{phiprime}
&&\varphi^\prime =\frac{\partial \varphi}{\partial
 H}=\frac{1}{8S}\ \frac{1}{\cos\varphi +
 D\sin\varphi\left(\displaystyle 1+\frac{1}{2\cos^{2}\varphi}\right)} \ 
,\nonumber\\ 
&&\mbox{and} \ \ \ \frac{\partial
 C_1}{\partial\varphi}=\frac{D}{\cos^2\varphi}\ ,
\end{eqnarray}
in which we used Eq. (\ref{phi1}) and an explicit expression for
$C_1=1+D\tan\varphi$ from Eq. (\ref{Cperp}). Note that the order of 
$\delta E$ is ${\cal O}(S)$ while $\delta S$ and $\delta M$ are ${\cal O}(1)$.
\begin{figure}[t]  
\includegraphics[width=8cm]{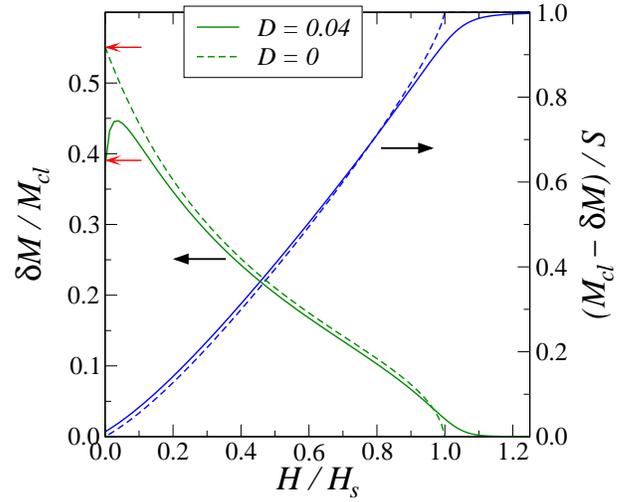}
\caption{(Color online) 
Uniform magnetization (classical part + $1/S$-correction),
  right axis,  
and the $1/S$ part to the uniform magnetization normalized to its
classical part, left axis, v.s. $H/H_s$. $D=0$ (dashed) and $D=0.04$ 
(solid). Arrows at the left axis show 
$\lim_{H\rightarrow 0}\lim_{D\rightarrow 0}$, Eq. (\ref{lim1}), and
$\lim_{D\rightarrow 0}\lim_{H\rightarrow 0}$, Eq. (\ref{lim2}). 
 $S=1/2$.}
\label{q_dM1}
\end{figure} 

{\it General features.} \ \ \ 
Having at hands expressions for the $1/S$ corrections to $E$, 
$\langle S^z\rangle$, and $M$, Eqs. (\ref{dEa}),
(\ref{dS}), and (\ref{dM}), together with the definitions of the
spin-wave energy, Eq. (\ref{wperp}), constants $C_i$,
Eq. (\ref{Cperp}), and the canting angle, Eq. (\ref{phi1}), one can
calculate their field-dependence.
Our Figure \ref{q_dE_dS} shows the results for $\delta
E/4S$ and $\delta S/S$ v.s. $H/H_s$, while 
Figure \ref{q_dM1} shows the field dependence of 
$M/S=(M_{cl}-\delta M)/S$ and $\delta M/M_{cl}$, all for 
$S=\frac12$ case. In both Figures the results for $D=0$ and $D=0.04$
are shown.

The overall field-dependence of the considered quantities is the same: 
quantum corrections are suppressed by the field.\cite{Zhitomirsky}
For the case of the pure Heisenberg model ($D=0$) the ground state
above the saturation field ($H\ge H_s$) is classical and all
quantum corrections vanish at $H=H_s$. Since in ${\bf H}\perp {\bf D}$
configuration the DM interaction prevents the full spin polarization
by the field, quantum effects survive in the region $H>H_s$. However,
with the further increase of the field they diminish quickly, see Figs.
\ref{q_dE_dS} and \ref{q_dM1}.

As is shown in Figures \ref{q_dE_dS} and \ref{q_dM1} different quantities 
are affected differently by the DM interaction. While we will elaborate on 
the detailed features of  such effects below, we would like to note that 
the quantum corrections to all the considered quantities are undoubtedly
suppressed by the DM interaction compared to the $D=0$ case for the fields
$H\alt H_s$.

Besides the proliferation of the quantum fluctuations into the $H>H_s$
region and an overall suppression of them for $H<H_s$ there are two
notable features shown in Figs. \ref{q_dE_dS} and \ref{q_dM1}. First, while
changes in the $1/S$-corrections to the energy due to a modest $D=0.04$ are 
hardly noticeable, changes in the quantum fluctuations of the on-site
magnetization $\delta S$ are substantial. Moreover, the relative role of such 
DM-induced changes seems to be significantly amplified by the applied field.
For instance, at $H=H_s/\sqrt{2}$  $\delta S$ is suppressed by almost 
40\% compared to its $D=0$ value due to the DM-term of the magnitude
only 4\% of $J$, see Fig. \ref{q_dE_dS}. Second, the behavior of 
$\delta M/M_{cl}$ v.s. $H$ is non-monotonic with significantly 
different zero-field limit of this quantity for the $D=0$ and $D\neq 0$
cases, see Fig. \ref{q_dM1}. 

{\it Uniform magnetization, $H=0$.}\ \ \  We would like to discuss the 
zero-field behavior of $\delta M/M_{cl}$ first. The reason for the choice of
such a quantity is simple: the ratio of the quantum $1/S$ part, $\delta M$, to 
the classical magnetization, $M=S\sin\varphi$, shows explicitly the
relative role of the quantum correction in the uniform magnetization.
In addition,
in the pure Heisenberg $D=0$ case all the terms 
in the $1/S$-expansion of the uniform magnetization vanish identically
as $H\rightarrow 0$.  Thus, at $D=0$ and
$H\rightarrow 0$\  $\delta M/M_{cl}$ is also related to the 
$1/S$-correction to the transverse susceptibility: $\chi=\chi_{cl}-
\delta\chi=(M_{cl}-\delta M)/H$ [recall that $\sin\varphi=H/H_s$ when
$D=0$]. Thus, using Eq. (\ref{dM}) and after 
some algebra the zero-field value of $\delta M/M_{cl}$ for $D=0$ case 
can be written as:
\begin{eqnarray}
\label{lim1}
\lim_{H\rightarrow 0}\lim_{D\rightarrow 0}\frac{\delta M}{M_{cl}}=
\frac{1}{2S}\sum_{\bf k}\gamma_{\bf k}\sqrt{\frac{1+\gamma_{\bf k}}
{1-\gamma_{\bf k}}}=\frac{0.55115}{2S} , 
\end{eqnarray}
compare with the expression for $\delta\chi$ in Ref. 
\onlinecite{Zhitomirsky}. This numerical value is marked
by the upper arrow pointing towards the left $y$-axis in
Fig. \ref{q_dM1}. One can see, however, that the DM interaction changes 
this quantity drastically. Neither the classical nor the quantum part of the 
uniform magnetization vanish in zero field anymore because of the weak 
ferromagnetism induced by the DM term. In other words, canting angle $\varphi$
does not go to zero at $H\rightarrow 0$, but approaches a finite value
$\simeq D/2$. 
This leads to the following expression for the opposite order of limits: 
\begin{eqnarray}
\label{lim2}
\lim_{D\rightarrow 0}\lim_{H\rightarrow 0}\frac{\delta M}{M_{cl}}=
\frac{1}{2S}\sum_{\bf k}\left(
\frac{1}{\sqrt{1-\gamma^2_{\bf k}}}-1\right)=\frac{0.3932}{2S} , 
\end{eqnarray}
which is nothing but the expression for the $1/S$-correction to the 
on-site (staggered) magnetization $\delta S/S$ in the pure Heisenberg
model at $H=0$. This value is marked by the lower arrow pointing 
towards the left $y$-axis in Fig. \ref{q_dM1}. Clearly,
the behavior of $\delta M/M_{cl}$ v.s. $H$ and $D$ is singular 
as the limits $H\rightarrow 0$ and $D\rightarrow 0$ taken in different
order lead to different results. For $D\neq 0$ there is a maximum in
$\delta M/M_{cl}$ v.s. $H$ around $H\simeq D/2$ and then this quantity
decreases roughly parallel to its $D=0$ counterpart, see Fig. \ref{q_dM1}.
\begin{figure*}
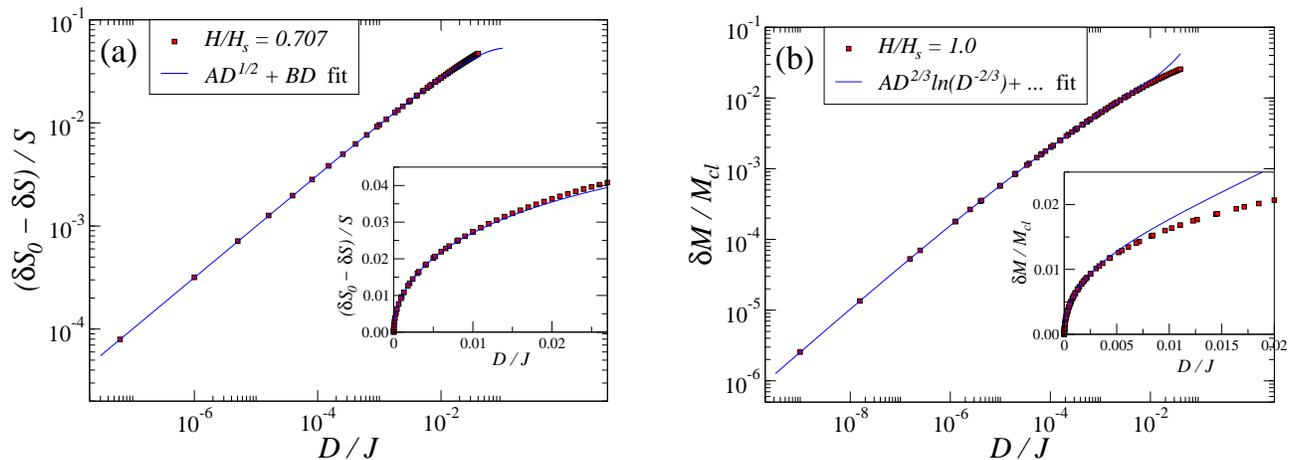
 
\includegraphics[width=8cm]{q_dS_vs_A_Pi_4}\hskip 1cm
\includegraphics[width=8cm]{q_dM_vs_A}
\caption{(Color online) 
The DM-induced changes in the $1/S$-corrections to (a) the on-site 
magnetization at $H=H_s/\sqrt{2}$ and (b) the uniform magnetization at $H=H_s$.
Exact results obtained by using the numerical integration in (a) Eq. (\ref{dS})
and (b) Eq. (\ref{dM}) (squares) and the asymptotic expressions 
(a) Eq. (\ref{dSasym}) and (b) Eq. (\ref{dMasym}) (solid lines) are shown. (a)
$A=1/2\pi S$, $B=0.48$, (b) $A=1/3\pi S$. $S=1/2$.}
\label{q_dM_dS_vs_A}
\end{figure*} 

{\it On-site magnetization.} \ \ \ To address the question on the 
sensitivity of the quantum component of the on-site magnetization,
$\delta S$, to the DM interaction it is useful to analyze the quantity
$\Delta S^{DM}=\delta S_0-\delta S$ as a function of $D$, where the subscript 
refers to the $D=0$ value.
Using Eq. (\ref{dS}) one can write:
\begin{eqnarray}
\label{ddS}
\frac{\Delta S^{DM}}{S}=\frac{\delta S_0-\delta
  S}{S}=\frac{1}{2S}\sum_{\bf k}\left( 
\frac{A_{\bf k}^0}{\nu_{\bf k}^0}-\frac{A_{\bf k}}{\nu_{\bf k}}
\right)\ ,
\end{eqnarray}
where $A_{\bf k}=C_1+C_2\gamma_{\bf k}$ with $C_1$ and $C_2$ from
Eq. (\ref{Cperp}) 
and $\nu_{\bf k}=\omega^\perp_{\bf k}/4S$ from Eq. (\ref{wperp})
as before, and $\nu_{\bf k}^0$ and $A_{\bf k}^0$ are their
$D=0$ limits, Eq. (\ref{w0}).
One can show that in the limit $D\ll 1$ the difference
 $\delta S_0-\delta S$ is
dominated by the contributions from the region in ${\bf k}$-space
close to the $(\pi,\pi)$-point with the rest of the 
Brillouin zone contributing to the subleading terms only. 
It may thus be concluded that the behavior of the DM-induced change in 
$\delta S$ should be related to the DM-induced $(\pi,\pi)$-gap discussed in 
Sec. \ref{gaps}.2. After some algebra one can obtain the asymptotic 
expressions:
\begin{eqnarray}
\label{dSasym}
\frac{\Delta S^{DM}}{S}\approx\frac{1}{2\pi S}
\left\{
\begin{array}{ll}
D,&\ \  H=0\ ,\\ \\
\sqrt{D\sin 2\varphi},& \ \  D\ll H\alt H_s \ ,\\ \\
(\sqrt{3}-2) D^{2/3},&\ \  H = H_s\ ,
\end{array}
\right. 
\end{eqnarray}
where $\sin \varphi \simeq H/H_s$. Thus, aside from a somewhat 
more subtle numerical coefficient for the result at $H=H_s$, 
the considered quantity follows almost identically the 
$D$-dependence of the $(\pi,\pi)$ gap, compare with 
Eqs. (\ref{gap2a}), (\ref{gap2b}). 
This asymptotic consideration also shows that except for the close vicinity
of $H=0$ and $H=H_s$ the DM-induced change in $\delta S$ is proportional to 
$\propto D^{1/2}$. That is, it is significantly amplified compared to the 
other quantities and is not small already for the moderate values of $D$ as we 
observed in Fig. \ref{q_dE_dS}. To demonstrate the validity of this asymptotic
result, Eq. (\ref{dSasym}), one can take the integral in
Eq. (\ref{ddS}) numerically 
and compare it with the results of Eq. (\ref{dSasym}) in the $D\ll 1$ limit. 
Such a comparison for a representative value of the field $H=H_s/\sqrt{2}$ 
($\sin 2\varphi =1$ in Eq. (\ref{dSasym})) is shown in
Fig. \ref{q_dM_dS_vs_A}(a) in both log-log and linear scales. The
coefficient $B=0.48$ for the subleading term  
(${\cal O}(D)$) is used in this plot. One can see an excellent agreement
of the numerical integration of an exact expression (squares) with the
asymptotic formula, Eq. (\ref{dSasym}), (solid lines).

{\it Uniform magnetization.} \ \ \ A similar analysis can be given to
the other  
quantities. We have already discussed above the DM-induced changes in
$\delta M/M_{cl}$ 
in the zero-field limit. Introducing $\Delta M^{DM}=\delta M_0-\delta M$ to 
characterize the effect of the DM interaction on the quantum component
of the uniform magnetization, where $\delta M_0$ is the $D=0$ limit of
$\delta M$, 
after some algebra one obtains the asymptotic expressions: 
\begin{eqnarray}
\label{dMasym}
\frac{\Delta M^{DM}}{M_{cl}}\approx\frac{1}{2S}
\left\{
\begin{array}{ll}
0.1574,& \  H=0\ ,\\ \\
A_1 D,&  \  D\ll H\alt H_s \ ,\\ \\
\displaystyle
\frac{2D^{2/3}}{3\pi}\ln D^{2/3},&
\   H = H_s\ .
\end{array}
\right. 
\end{eqnarray}
Clearly, these are less trivial dependencies than for the on-site
magnetization,  
Eq. (\ref{dSasym}). The number for the $H=0$ case is simply the difference 
of the results in Eqs. (\ref{lim1}) and (\ref{lim2}) discussed before.
For the intermediate-field regime $D\ll H\alt H_s$ the DM-induced
changes in the uniform magnetization are {\it not} dominated by any particular 
region in the ${\bf k}$-space, but are accumulated over the entire BZ. 
Therefore, there is no straightforward relation between the $D$- or
$H$-dependence 
of the $(\pi,\pi)$-gap and $\Delta M^{DM}$. The expression for $A_1(H)$
is rather cumbersome and we only list its value for a representative field:
$A_1(H_s/\sqrt{2})=0.1580$. On the contrary, in the regime $H=H_s$ the
DM-induced 
changes in $\delta M/M_{cl}$ {\it are} dominated by the vicinity of the 
$(\pi,\pi)$-gap, but only with the logarithmic accuracy, see
Eq. (\ref{dMasym}), since the subleading term is ${\cal O}(D^{2/3})$. 
To demonstrate the validity of the latter asymptotic
expression we evaluated $(\delta M_0-\delta M)/M_{cl}$ numerically
using Eq. (\ref{dM}) 
and compared it with the results in Eq. (\ref{dMasym}) for $H=H_s$.
An excellent agreement between the numerical (squares) and asymptotic
expressions (solid lines) is shown in Fig. \ref{q_dM_dS_vs_A}(b).

{\it Energy.} \ \ \ A similar consideration can be given to the
$1/S$-correction to the energy. For the intermediate-field regime
$D\ll H\alt H_s$  one finds that $(\delta E_0-\delta E)/4S=A_3 D+{\cal
  O}(D^{3/2})$, same as $\delta M^{DM}$ but with much smaller
numerical coefficient.  
For example, for $H=H_s/\sqrt{2}$ the expression for the coefficient
$A_3$ can be simplified to
\begin{eqnarray}
\label{A3}
A_3=\frac12\sum_{\bf k}\left(\frac{1-\gamma_{\bf
    k}^2/2}{\sqrt{1+\gamma_{\bf k}}} -1\right)=0.0259 . 
\end{eqnarray}
 At the saturation field the energy correction can be approximately written as:
\begin{eqnarray}
\label{dE1}
\frac{\delta E}{4S}=
\frac{D^{4/3}}{4\pi}\ln \left(\frac{D^{2/3}}{\lambda}\right)+{\cal O}(D^2), 
\end{eqnarray}
which contains an extra power of $D^{2/3}$ in comparison with $\delta
M/M_{cl}$,  Eq. (\ref{dMasym}).

{\it Summary.}\ \ \ Altogether, the quantum $1/S$ corrections are
changed substantially 
under the influence of the DM interaction. Many of the considered
changes depend 
on $D$ non-analytically, see Eqs. (\ref{dSasym}), (\ref{dMasym}), and
(\ref{dE1}), and some  
exhibit a singular behavior as for the case of $\delta M/M_{cl}$ in
$(H, D)\rightarrow 0$  
limit, see Eqs. (\ref{lim1}), (\ref{lim2}), and Fig. \ref{q_dM1}. The most 
substantial change is seen in the on-site magnetization where for the
broad field range 
$\delta S_0-\delta S\propto D^{1/2}$, see Eq. (\ref{dSasym}) and
Fig. \ref{q_dE_dS}.  
Except for the vicinity of the saturation field where the quantum
effects are enhanced (or, rather, induced for $H>H_s$), 
the quantum fluctuations are suppressed by the
DM interaction, see Figs. \ref{q_dE_dS} and \ref{q_dM1}.

\section{K$_2$V$_3$O$_8$}
\label{KVO}
\subsection{Experiments and Hamiltonian}

K$_2$V$_3$O$_8$ has been studied experimentally by magnetometry 
and neutron scattering, as reported in Ref. \onlinecite{Lumsden}.
 This material contains weakly coupled square lattice 
planes of spins $S=\frac12$. The spins interact antiferromagnetically with the
superexchange constant $J=12.6$K. In the ground state spins are
pointing along the  
$z$-axis while in a low external field ($H<1$T) two types of
transitions 
were seen. For the field along the $z$-axis the well-known
``spin-flop''-type transition occurs at $H_{SF}$: spins 
suddenly ``flop'' in the $x-y$ plane and cant towards the field upon 
further increase of the field in a configuration similar to 
Fig. \ref{angles}(a). For the field ${\bf H}\perp z$ an unusual 
``spin-rotation'' transition is seen: spins perform a rotation by
$\pi/2$ angle gradually from along the $z$-axis into the $x-y$ 
plane within a very small field 
range $0<H<H_{SR}$. Above the ``spin-rotation'' field $H_{SR}$ spins 
are in the $x-y$ plane and cant towards the field in a configuration 
similar to Fig. \ref{angles}(b). Such a behavior made the authors of 
Ref. \onlinecite{Lumsden} to conclude that there are small
anisotropies 
of two types which accompany the superexchange in this material: 
the easy-axis anisotropy along the $z$-axis and the DM-anisotropy 
also directed along the $z$-axis. Therefore, K$_2$V$_3$O$_8$  is
 described by our Hamiltonian (\ref{H}) with an additional easy-axis term:
\begin{eqnarray}
\label{dH}
\delta{\cal H}_{ea}= E\sum_{\langle ij \rangle}S^z_i S^z_j \ .
\end{eqnarray}
As is shown in Ref. \onlinecite{Lumsden}, with an appropriate choice 
of parameters the Hamiltonian (\ref{H})$+$(\ref{dH}) describes
excellently all the observed transitions.

As is mentioned in Sec. \ref{Q_effects}, although the easy-axis
constant $E$ is of the order of ${\cal O}(D^2)$, it competes with the
DM-term on equal footing. While for an ideal staggered configuration
of the DM vectors it is expected that the easy-axis constant is
$E=D^2/2$ and the system possesses a hidden $O(3)$
symmetry,\cite{Shekhtman1} a non-staggered component of the DM
vector generally 
breaks such a high symmetry and leads to $E< D^2/2$.\cite{Yildrim}
In realistic systems, like K$_2$V$_3$O$_8$ the easy-axis term is, de
facto, $E>D^2/2$ due to some other microscopic processes. 
We thus consider $D$ and $E$ as independent parameters, but still
$E\sim D^2$. In the rest of the Section we consider the experimentally
relevant case $E>D^2/2$ and discuss the case $E<D^2/2$ in the end.

One can show that for the case of vector ${\bf D}$ being along the 
``easy'' axis and for 
$E>D^2/2$ (we continue to use $J=1$ units here), the easy-axis anisotropy 
overcomes the DM-term in the ground state. That is, instead of staying 
in the plane $\perp {\bf D}$ and canting towards each other, as
favored 
by the DM interaction, spins align along the ``easy'' axis as dictated 
by the easy-axis term and as if the DM-term was absent. Therefore, 
in the ground state of this Hamiltonian at $H=0$ and $E>D^2/2$ there 
is no weak ferromagnetism and the DM interaction is
``hidden''. However, 
the external field helps to reveal the presence of the DM interaction
in the system. It is the interplay of the easy-axis and the DM-term 
which causes an unusual spin rotation from the $z$-axis into the $x-y$ 
plane in K$_2$V$_3$O$_8$.

\subsection{Ground state and excitation spectrum in a field}

We thus consider the Hamiltonian (\ref{H})$+$(\ref{dH}) as a starting 
point and take the experimental estimate for the DM constant 
$D=0.04$, Ref. \onlinecite{Lumsden}. For the easy-axis constant 
we use the value $E=0.0017$, somewhat higher than estimated in
Ref. \onlinecite{Lumsden}, for the reasons discussed below.  

Before the technical discussion we would like to describe the field 
evolution of the ground state from the symmetry point of view. In zero 
field for $H<H_{SF} (H_{SR})$ the continuous 
$O(3)$ symmetry is broken by the easy-axis term to a discrete 
Ising-like one.
Thus, all the excitations in such a phase should be gapped. 
For ${\bf H}\| z$ above the 
spin-flop field, $H>H_{SF}$, the continuous symmetry is restored to 
$O(2)$. This symmetry corresponds to the freedom of choice of the direction 
of the spins' component in the $x-y$ plane. Since the DM vector 
is also $\| z$, its presence does not change this consideration 
except that the spins acquire a small canting angle towards each other
in the $x-y$ plane, 
see Fig. \ref{angles}(a). Therefore, one of the magnon branches 
should be gapless in this phase. This is identical to the 
${\bf H}\|{\bf D}$ field configuration discussed in Secs. 
\ref{class}.1, \ref{gaps}.1 without the easy-axis anisotropy. For 
the in-plane field ${\bf H}\perp {\bf D}\| z$ above the spin-rotation 
field, $H>H_{SR}$, all the continuous symmetries remain broken and the 
ground state is identical to the one considered in Sec. \ref{class}.2 
and Fig. \ref{angles}(b) where spins are in the 
$x-y$ plane and are canted towards the field. In such a phase no 
gapless excitation should exist.

With these ideas in mind we will analyze classical spin 
configurations and excitation spectrum of the Hamiltonian 
(\ref{H})$+$(\ref{dH}). Because of the easy-axis term there 
are two different phases for each of the field directions: 
below and above the spin-flop 
(spin-rotation) transition for ${\bf H}\| z$ (${\bf H}\perp z$).

\subsubsection{Out-of-plane field.}
\begin{figure}[t]  
\includegraphics[width=8cm]{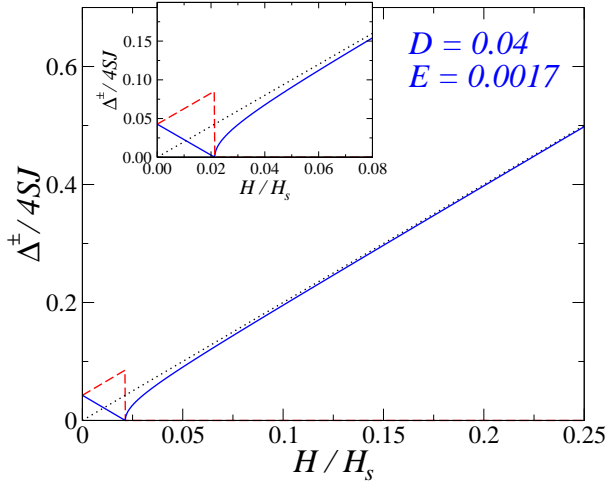}
\caption{(Color online) 
The field-evolution of the gaps for the spin-flop field
  configuration ${\bf H}\|{\bf D}\|z$. The spin-flop transition occurs
  at $H_{SF}$, Eq. (\ref{HSF}). Below the transition the gaps depend
  linearly on $H$, Eq. (\ref{dpm}). 
Above the transition one of the gaps is zero
  (dashed) and
  the other corresponds to a uniform precession of the field-induced
  magnetization (solid), Eq. (\ref{d00}). 
The uniform precession mode for the $E=D=0$
  case is shown by the dotted line ($\Delta =H$). Inset: same results
  in the region of the small fields. Saturation field for
  K$_2$V$_3$O$_8$ is $H_s=8SJ=37.6$T.}
\label{H||D}
\end{figure} 

For the ${\bf H}\|{\bf D}\|z$ configuration the classical energy of the
Hamiltonian (\ref{H})$+$(\ref{dH}) can be written as:
\begin{eqnarray}
\label{E3}
&&\frac{E_{cl}}{2NS^2}=\cos\theta_A\cos\theta_B(1+E)\nonumber\\
&&\phantom{\frac{E_{cl}}{2NS^2}=}-\sin\theta_A\sin\theta_B
\left(\cos2\varphi+D\sin 2\varphi\right)\\
&&\phantom{\frac{E_{cl}}{2NS^2}=}
-\frac{H}{4S}\left(\cos\theta_A+\cos\theta_B\right)\ ,\nonumber
\end{eqnarray}
where $\theta_A$ and $\theta_B$ are the angles with the $z$-axis made
by spins in the $A$ and $B$ sublattice, respectively. $\varphi$ is the
angle in the $x-y$ plane as in Fig. \ref{angles}(a). Using the symmetry
arguments one can show that for the case $E>D^2/2$ two choices of the
spin configuration are possible: (i) $\theta_A=0$, $\theta_B=\pi$, and 
(ii) $\theta_A=\theta_B=\theta$, as in Fig. \ref{angles}(a). Needless
to say, these two choices correspond to $H<H_{SF}$ and $H>H_{SF}$,
respectively. For the first choice the spins are pointing along the
$z$-axis and the classical energy is field- and DM-independent:
\begin{eqnarray}
\label{E4}
\frac{E_{cl}^{(1)}}{2NS^2}=-(1+E)\ .
\end{eqnarray}
For the second choice the energy is minimized by the following choice
of angles:
\begin{eqnarray}
\label{phi_theta}
\varphi=\frac12\tan^{-1}D\ ; \ \  \ 
\cos\theta=\frac{H}{8S}\ \frac{2}{1+E+\sqrt{1+D^2}}\ , 
\end{eqnarray}
where $\varphi$ gives a small field-independent canting of spins
in the $x-y$ plane. 
For small fields spins are almost in the $x-y$
plane as $\theta\approx\pi/2$, all 
in a very close similarity with our discussion of the DM-only case,
Sec. \ref{class}.1, Fig. \ref{angles}(a), and Eq. (\ref{theta}). Using
these angles (\ref{phi_theta}) one can rewrite the classical energy as:
\begin{eqnarray}
\label{E5}
\frac{E_{cl}^{(2)}}{2NS^2}=-\sqrt{1+D^2}-
\frac{H^2}{16S^2(1+E+\sqrt{1+D^2})}\ .
\end{eqnarray}
From the condition $E_{cl}^{(1)}=E_{cl}^{(2)}$ 
we can now find the field at which the spin-flop transition between
the two spin configurations occurs. This gives:
\begin{eqnarray}
\label{HSF}
H_{SF}=4S\sqrt{2E-D^2+E^2}\simeq 4S\sqrt{2E-D^2} \ ,
\end{eqnarray}
which is well-defined only if $E>D^2/2$.
Because $E\sim D^2\ll 1$ one can neglect $E^2$, $D^4$, etc. as being
extremely small. 
One can verify that the first derivative of the energy with respect 
to field has a jump as is expected for the first-order transition. 
The angles also experience
discontinuous jumps to finite values at $H_{SF}$ as can be inferred
from Eq. (\ref{phi_theta}).

The field evolution of the excitation spectrum within the spin-flop
problem has been studied in the past and is well-known (see, e.g., Ref.
\onlinecite{Tennant}). In our case there are some quantitative 
changes due to the DM-term, but the overall behavior is very similar.
Below the transition there are two branches:
\begin{eqnarray}
\label{wkpm}
\omega_{\bf k}^\pm =4S\sqrt{(1+E)^2-\gamma_{\bf k}^2(1+D^2)}\pm H\ ,
\end{eqnarray}
with the gaps:
\begin{eqnarray}
\label{dpm}
\Delta^\pm =H_{SF}\pm H\ ,
\end{eqnarray}
where the spin-flop field is given in Eq. (\ref{HSF}).

Above the spin-flop transition the spectrum can be written as:
\begin{eqnarray}
\label{wkpm1}
&&\omega_{\bf k} =4S\sqrt{1+D^2}\sqrt{(1+\gamma_{\bf
    k})(1-\delta_1\gamma_{\bf k})}\ ,\\
&&\mbox{where} \ \delta_1=
\left(1+E-\frac{4(H/H_s)^2}{1+E+\sqrt{1+D^2}}\right)
\frac{1}{\sqrt{1+D^2}}\ , \nonumber
\end{eqnarray}
with one mode gapless, $\Delta_{\pi\pi}\equiv 0$, and the other mode 
having a gap which can be written as:
\begin{eqnarray}
\label{d00}
\Delta_{00} =
\sqrt{\frac{2\sqrt{1+D^2}}{1+E+\sqrt{1+D^2}}}\ \sqrt{H^2-H_{SF}^2}
\ .
\end{eqnarray}
This mode reaches the uniform precession behavior $\Delta\simeq H$ 
at $H\gg H_{SF}$. All these field-dependencies of the gaps are shown
in our Fig. \ref{H||D}.
One can see a jump in one of the gaps associated with the spin-flop
transition. The dotted line is $\Delta=H$ ($E=D=0$ result). As we
discussed above, one of the modes is necessarily gapless for
$H>H_{SF}$, same as for the $E=0$ case, Sec. \ref{gaps}.1.
\begin{figure}[t]  
\includegraphics[width=8cm]{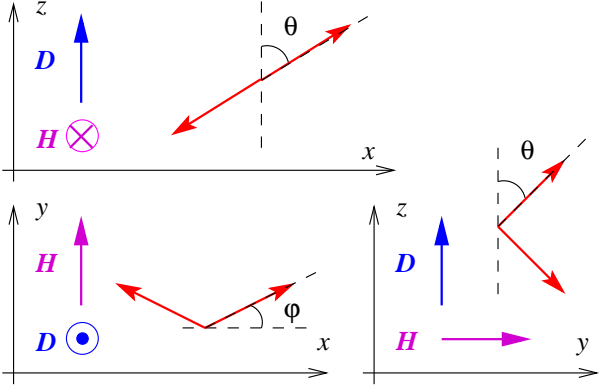}
\caption{(Color online) 
The projections of the spin arrangement on the $x-z$-,
  $x-y$-, and  $y-z$-planes within the spin-rotation phase
  $0<H<H_{SR}$. 
While $\varphi$ is the angle 
between the {\it projections} of spins on the $x-y$ plane and the
$x$-axis, $\theta$ is the angle between the actual spin
direction and the $z$-axis. ${\bf H}$ is chosen along the $y$-axis.}
\label{angles1}
\end{figure} 

\subsubsection{In-plane field.}

For the ${\bf H}\perp{\bf D}\|z$ field configuration
the spin arrangement within the spin-rotation phase $0<H<H_{SR}$ 
is given in Fig. \ref{angles1}. Note that $\varphi$ is the angle 
between the {\it projections} of spins on the $x-y$ plane and the
$x$-axis, while $\theta$ is the angle between the actual spin
direction and the $z$-axis. 
One can see that the angle with the $z$-axis
$\theta_B=\pi-\theta_A$ for the spins in different sublattices.
Both angles, $\varphi$ and $\theta$, are coupled non-trivially 
in the classical energy of the
Hamiltonian (\ref{H})$+$(\ref{dH}) which can be written as:
\begin{eqnarray}
\label{E6}
&&\frac{E_{cl}}{2NS^2}=-\cos^2\theta(1+E)\\
&&\phantom{\frac{E_{cl}}{2NS^2}=}
-\sin^2\theta\left(\cos2\varphi+D\sin 2\varphi\right)
-\frac{H}{2S}\sin\theta\sin\varphi\ .\nonumber
\end{eqnarray}
Minimizing the energy, after some algebra, leads to the following
expressions for the angles:
\begin{eqnarray}
\label{phi_theta1}
\tan\varphi=\frac{E}{D}\ ; \ \  \ 
\sin\theta=\frac{H}{4S}\ \frac{\sqrt{D^2+E^2}}{2E-D^2+E^2}\ , 
\end{eqnarray}
Somewhat surprisingly, the $\varphi$ angle remains  field-independent
throughout the spin-rotation phase. 
Since the easy-axis interaction $E$ is of the
same order as $D^2\ll 1$, $\varphi\approx E/D$ is also small. On the
contrary, $\theta$ changes drastically (but continuously) from $0$ at
$H=0$ to $\pi/2$ at the critical field $H_{SR}$. The latter can be found 
from Eq. (\ref{phi_theta1}) as:
\begin{eqnarray}
\label{HSR}
H_{SR}=4S\ \frac{2E-D^2}{D}\ ,
\end{eqnarray}
where we omitted ${\cal O}(E^2)$ terms.

For the fields above the transition, $H>H_{SR}$, a different solution
minimizes the classical energy in Eq. (\ref{E6}): $\theta=\pi/2$ is
constant and $\varphi$ is field-dependent. That is, the spins are
lying in the $x-y$ plane as depicted in Fig. \ref{angles}(b). It is
easy to see
that for $\theta=\pi/2$ Eq. (\ref{E6}) reduces to the classical
energy for the DM-only case, Eq. (\ref{E2}). Thus, the easy-axis term
is effectively ``switched-off'' by the field for $H>H_{SR}$. 
Because of that the
field-dependence of the canting angle $\varphi$ also remains the same as in
Sec. \ref{class}.2, Eq. (\ref{phi1}).

Using the canting angles, Eqs. (\ref{phi_theta1}) and
(\ref{phi1}), one can simplify the 
expressions for the classical energies below and above the
spin-rotation transition to:
\begin{eqnarray}
\label{E7a}
&&\frac{E^<_{cl}}{2NS^2}=-(1+E)-\frac{H^2}{16S^2}\ \frac{E}{2E-D^2}\ ,\\
\label{E7b}
&&\frac{E^>_{cl}}{2NS^2}=-1-\frac12\left(\frac{H}{4S}+
D\right)^2\ ,
\end{eqnarray}
where ${\cal O}(E^2)$ terms are neglected.
From these formulae one can verify that the spin-rotation
transition is of the second order as the energy and its first
derivative are continuous through the transition. The angles also
change continuously at $H_{SR}$ as can be seen, for instance, for
$E, D^2 \ll 1$:  $\varphi^<= E/D$ matches with $\varphi^>=D/2+H/8S$
at $H=H_{SR}$, Eq. (\ref{HSR}). 

These canting angles can be translated directly into the field-induced 
uniform magnetization which has been measured as a function
of applied field, see Ref. \onlinecite{Lumsden}. In small field the
uniform magnetization depends linearly on field,
but has different slopes for $H<H_{SR}$ and
$H>H_{SR}$. On the other hand, such slopes can be extracted from our
Eqs. (\ref{E7a}) and (\ref{E7b}) which give:
\begin{eqnarray}
\label{MM}
\left(\frac{dM^>}{dH}\right)\bigg/
\left(\frac{dM^<}{dH}\right)=1-\frac{D^2}{2E}\ .
\end{eqnarray}
This relation is useful to extract the ratio $D^2/E$ because the
slope in the magnetization v.s. field can be measured rather reliably
and is the subject of less experimental error than, say, the transition
field itself. It is using this relation and $D=0.04$ we have
extracted the value $E=0.0017$ for K$_2$V$_3$O$_8$ used throughout
this Section.

After the easy-axis and the DM constants are chosen one can find 
the $T=0$ values of the spin-flop and spin-rotation fields from Eqs. 
(\ref{HSF}) and (\ref{HSR}). Using the values of $g_c=1.922$ and 
$g_{ab}=1.972$ for K$_2$V$_3$O$_8$ reported in Ref. \onlinecite{Lumsden} 
one finds:
$H_{SF}\approx 0.85$T and $H_{SR}\approx 0.89$T. These theoretical 
values 
are within the same range as  $H_{SF}^{exp}=0.85$T and $H_{SR}^{exp}= 0.65$T
found in Ref. \onlinecite{Lumsden} from the neutron scattering and 
magnetization measurements and $H_{SF}^{exp}=0.95$T and $H_{SR}^{exp}= 0.55$T
estimated from the upturn in the thermal conductivity v.s. field in Ref. 
\onlinecite{Sales}. Since all these experimental results were obtained at 
a finite temperature of the order of the N\'eel ordering temperature for 
this material, the differences between the theoretical results and the data 
are to be expected. The closer agreement with the data for the spin-flop 
field are due to the first-order nature of that transition. 
It is known\cite{Tennant} that the critical field for the spin-flop 
transition is only weakly temperature-dependent. For the 
spin-rotation transition, on the other hand, one can expect a 
suppression of the critical field with the temperature. Overall, the
theoretical results for the critical fields are in a close accord with 
the experimental data. 

One can now address the field-dependence of the 
excitation spectrum for the spin-rotation
problem. As is explained in Sec. \ref{gaps}, we need the
field-dependence of  the canting
angles, a transformation from the local to the laboratory frame for
spins, and the subsequent quantization of the spin Hamiltonian to obtain
the spectrum. While above the transition, $H>H_{SR}$, 
the first two steps are essentially identical to the ones performed
for the DM-only case in Sec. \ref{gaps}.2, below the transition,
$H<H_{SR}$, this procedure requires an additional and a somewhat 
cumbersome algebra. We thus simply provide here the results for the gaps
in the spin-rotation phase $H<H_{SR}$:
\begin{eqnarray}
\label{dd<a}
&&\Delta_{00}=\sqrt{4SDH_{SR}\left(1+\frac{HH_{SR}}{4SD}\right)}\ ,\\
\label{dd<b}
&&\Delta_{\pi\pi}=\sqrt{4SDH_{SR}\left(1-\frac{H^2}{H_{SR}^2}\right)}\ ,
\end{eqnarray}
where $H_{SR}$ is from Eq. (\ref{HSR}) and we 
neglect ${\cal O}(E^2)$ terms as before.
\begin{figure}[t]  
\includegraphics[width=8cm]{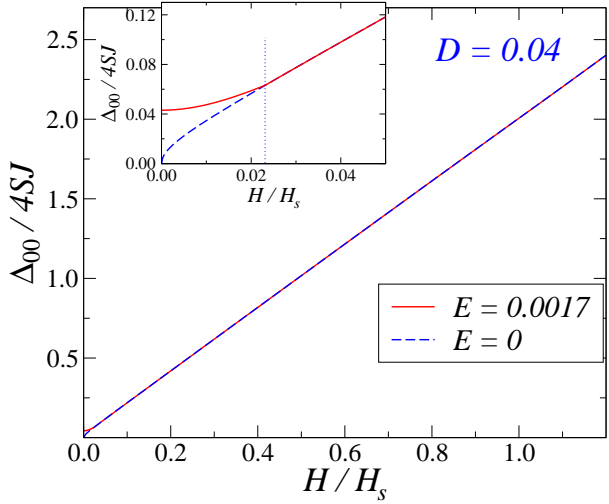}
\caption{(Color online) 
$\Delta_{00}/4SJ$ v.s. $H/H_s$ for the in-plane field
 direction, $D=0.004$, and
 two values of $E$: $E=0.0017$ (solid), Eqs. (\ref{dd<a}) and
 (\ref{dd>a}), 
and $E=0$ (dashed), Eq. (\ref{gap1}). 
Inset: same plot in the region of small fields. The
 spin-rotation transition is marked by the dotted vertical line.
Saturation field for
  K$_2$V$_3$O$_8$ is $H_s=8SJ=37.6$T.}
\label{gapsAB00}
\end{figure} 

For the fields above the transition, $H>H_{SR}$, the changes with
respect to the DM-only case, Sec. \ref{gaps}.2, 
are rather incremental and we can afford
listing the parameters of the spin-wave Hamiltonian:
\begin{eqnarray}
\label{Cperp1}
&&C_1=1+D\tan\varphi\ , \nonumber\\
&&C_2+C_3=-\cos 2\varphi - D\sin 2\varphi\ ,\\
&&C_2-C_3=1+B\ , \nonumber
\end{eqnarray}
which one can compare to an equivalent expression in 
Eq. (\ref{Cperp}). Thus, the easy-axis term leads to rather
obvious changes in the spin-wave energy, Eq. (\ref{wperp}). The gaps
are changed accordingly and are given by:
\begin{eqnarray}
\label{dd>a}
&&\Delta_{00}=4S\sqrt{\frac{H}{2S}\sin\varphi
\left(1+\frac12D\tan\varphi+E\right)} \ ,\\
\label{dd>b}
&&\Delta_{\pi\pi}=4S\sqrt{2(D\tan\varphi-E)}\\
&&\phantom{\Delta_{\pi\pi}=4S}
\times\sqrt{\left(1+D\tan\varphi-\frac{H}{8S}\sin\varphi\right)}\nonumber \ ,
\end{eqnarray}
which should be compared to the DM-only results, Eqs. (\ref{gap1}) and
(\ref{gap2}), respectively. It is important to note that  
in the vicinity of the transition field the
$(\pi,\pi)$-gap has the following non-analytic field-dependence:
\begin{eqnarray}
\label{ddapp}
\Delta_{\pi\pi}=\sqrt{4SD(H-H_{SR})}\ ,
\end{eqnarray}
for $H>H_{SR}$. This is reminiscent of Eq. (\ref{gap2a}) for the $H\ll H_s$
case, but with the field ``shifted down'' by  $H_{SR}+4SD$.

\begin{figure}[t]  
\includegraphics[width=8cm]{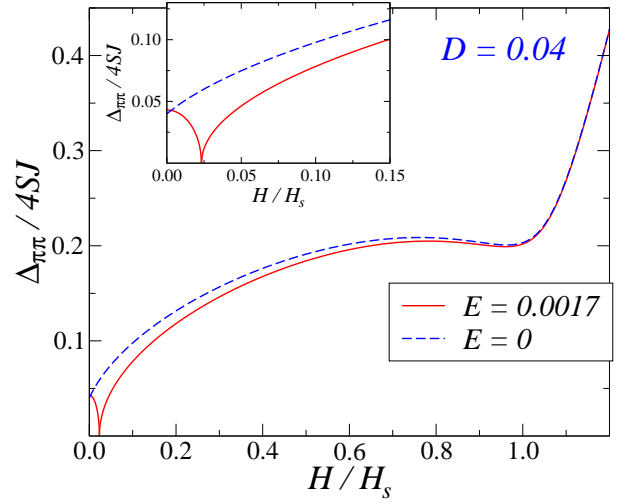}
\caption{(Color online) 
$\Delta_{\pi\pi}/4SJ$ v.s. $H/H_s$ for the in-plane field
 direction, $D=0.004$, and
 two values of $E$: $E=0.0017$ (solid), Eqs. (\ref{dd<b}) and
 (\ref{dd>b}), 
and $E=0$ (dashed), Eq. (\ref{gap2}). 
Inset: same plot in the region of small fields. Saturation field for
  K$_2$V$_3$O$_8$ is $H_s=8SJ=37.6$T.}
\label{gapsAB}
\end{figure}

Our results for the field-dependence of the $\Delta_{00}$ and
$\Delta_{\pi\pi}$ gaps, both below and above
the spin-rotation transition, together with the results for the
DM-only ($E=0$) case are shown in Figures \ref{gapsAB00} and
\ref{gapsAB}, respectively.
One can see that for the $\Delta_{00}$ gap above the transition the
results for K$_2$V$_3$O$_8$ are essentially indistinguishable from the
DM-only case. It can be shown that the difference between them scales
as $\propto {\cal O}(EH)={\cal O}(D^2H)$. 
The changes in  $\Delta_{\pi\pi}$ with
respect to the DM-only case are more substantial. They scale as ${\cal
O}(E/D)$ for the fields $H\agt H_{SR}$ but diminish closer to the
saturation field. Thus, the major effect of the easy-axis anisotropy
on the excitation spectrum is the effective ``off-set'' of the
external field by $H_{SR}+4SD$, as can be seen from comparison of
Eqs. (\ref{ddapp}) and (\ref{gap2a}). 

\subsubsection{$E\le D^2/2$ case.}

For completeness, we briefly discuss here the case $E<D^2/2$. In such
a case the ground state and the spectrum field-dependencies are
qualitatively the same as in the DM-only case (\ref{H}) considered in
Secs. \ref{class} and \ref{gaps}. In fact, for the most interesting
${\bf H}\perp {\bf D}$ configuration the canting angle in the ground
state is simply identical to Eq. (\ref{phi1}) because the easy-axis
anisotropy is effectively switched-off. In particular, the zero-field
ground state for $E<D^2/2$ possesses the same uniform magnetization as
if the easy-axis term would be absent. Interestingly enough, the
high- and intermediate-field behavior of the $(\pi,\pi)$-gap also
depends only on the DM-term, $\Delta_{\pi\pi}\propto (DH)^{1/2}$ for
$H\gg D$. On the other hand, the zero-field gap in this case is reduced:
$\Delta_{\pi\pi}=4S\sqrt{D^2-2E}$. Thus, the zero-field canting angle
$\varphi\simeq D/2$ is not in one-to-one correspondence anymore with the
zero-field  $(\pi,\pi)$-gap, in contrast to the DM-only case,
model (\ref{H}). 

A particular case $E=D^2/2$ is also interesting as it shows no gap in
zero-field, yet a finite-field behavior of the gap is governed by the
DM-term: $\Delta_{\pi\pi}\propto (DH)^{1/2}$, the result obtained in
Refs. \onlinecite{Sato,Mila}. Another interesting feature of that case
is that the classical energy in zero-field, Eq. (\ref{E6}), is
degenerate for any choice of $\theta$, while
$\varphi=\frac12\tan^{-1}D\simeq D/2$. That is, the weak
ferromagnetism with the uniform magnetization of an arbitrary strength
from the range $0<M\alt SD/2$ can occur in zero-field ground state 
as a result of
the spontaneous choice of $\theta$ in the spin configuration depicted
in Fig. \ref{angles1}. This scenario has been also discussed on
Ref. \onlinecite{Shekhtman1}.

\subsubsection{Summary.}

Altogether, an additional easy-axis anisotropy introduces some new
transitions in the physics of a realistic system in comparison with
the DM-only case, but leaves most of
the results for the field above the corresponding transition
unchanged. That is, for the out-of-plane 
field configuration one of the
gaps in the spin-wave spectrum
behaves similarly to a uniform precession mode, $\Delta\approx H$, 
while the other mode stays gapless. For the in-plane 
field both modes are gapped.  One of them is the uniform
precession mode while the other depends non-analytically on the field
and shows a non-trivial behavior in high field. We would like to note
that some indications of such a drastically different behavior of the
gaps for
different field directions have been observed in the field-dependence
of the thermal conductivity\cite{future} in K$_2$V$_3$O$_8$, see
Ref. \onlinecite{Sales}. 
One of the crucial differences of the K$_2$V$_3$O$_8$ spectrum
field-dependence from the
idealized DM-only case, Eq. (\ref{H}), is the effective off-set of the 
field for the in-plane field direction, leading to the $\propto
\sqrt{D(H-H_{SR})}$ behavior of the gap. 
Thus, in a realistic system, 
the low-field behavior of the DM-induced gap is
more delicate than simple $\propto \sqrt{DH}$
suggested recently.\cite{Affleck,Sato,Mila}

\section{Conclusions}
\label{Conclusions}

We conclude by summarizing our results. 
We have studied the effects of external field on the properties 
of an ordered Heisenberg antiferromagnet with the DM
interaction. Utilizing the spin-wave theory we have calculated the 
 quantum correction to the energy, on-site
magnetization, and uniform magnetization v.s. $H$ and $D$. 
We have shown that the spin-wave excitations
exhibit an unusual field-evolution of the gaps which 
leads to various non-analytic
dependencies of the quantum corrections.
We have found that the gap in one of the magnon branches follows a
particularly interesting behavior v.s. field for a specific field direction: 
$\Delta_{\pi\pi}$ is $\propto D$, $\propto D^{1/2}$, or  $\propto
D^{2/3}$ depending on the field. A similar behavior is reflected 
in some of the quantum corrections as well. 
We have demonstrated that for the fields $0<H\alt H_s$ the
DM interaction suppresses quantum fluctuations. The on-site ordered
moment is affected most strongly by such a suppression.
It is only in the regime of the field close to the saturation 
where the DM term enhances quantum fluctuations, effectively leading
to a proliferation of the quantum effects into the classical saturated
phase. We have shown that some quantum corrections  are singular
in the limit of $H, D\rightarrow 0$.

We also have considered K$_2$V$_3$O$_8$ which is well described by
the spin-$\frac12$, Heisenberg AF on the 
two-dimensional square lattice, 
with the additional DM and easy-axis anisotropies.
 We have demonstrated that K$_2$V$_3$O$_8$ is an
excellent candidate for the observation of the unusual
field-dependence of the gaps, characteristic to the 2D and 3D AFs. In
fact, an additional easy-plane anisotropy makes
it potentially possible to observe a non-analytic $\Delta\propto
\sqrt{D(H-H_{SR})}$ behavior of the gap.


\begin{acknowledgments}
I would like to thank Mark Lumsden and Stephen Nagler for numerous
discussions, sharing their experimental data, and patience. 
I am grateful to F. Mila, H. Ronnow,  and M. Zhitomirsky
for useful conversations and O. Tchernyshyov for insightful comments
and fruitful debates which led to an improvement of this work.
This work was supported by DOE under grant DE-FG02-04ER46174, and by
the ACS Petroleum Research Fund. 

\end{acknowledgments}




\begin{thebibliography}{99}
\bibitem{footnote2} The case of Cs$_2$CuCl$_4$ is severely complicated
  by the field-dependence of the AF ordering vector.
\bibitem{footnote5} A consideration of such an instability in the
  presence of anisotropies will be given in a future work.
\bibitem{footnote6} The coefficient $c$ in the transverse field in
  Ref. \onlinecite{Mila} should be equal to $D/2$ in the original DM
  interaction. Thus, the ``critical'' value for the occurrence of the
  minimum $c_c\simeq 0.031$ of Ref. \onlinecite{Mila} is close to 
our spin-wave result
  $D_c/2\simeq 0.033$.
\bibitem{future} The theoretical explanation of the thermal
  conductivity behavior\cite{Sales} 
in K$_2$V$_3$O$_8$ will be given elsewhere.
\end{thebibliography}
\end{document}